\documentclass[11pt,pdftex,titlepage]{article}
\usepackage{epsf}
\usepackage{epsfig}
\usepackage{graphicx}
\usepackage{times}
\usepackage{ifthen}
\usepackage{array}
\usepackage{comment}
\usepackage{url}
\usepackage{cellspace}
\usepackage{latexsym}
\usepackage[margin=1in]{geometry}
\usepackage{geometry}
\usepackage{amsfonts}
\usepackage{subfigure}
\usepackage{amsmath}
\usepackage{amsfonts}
\usepackage{multirow}
\usepackage{multicol}
\usepackage{tabularx}
\usepackage[table,xcdraw]{xcolor}
\usepackage{floatflt}
\usepackage{datetime}
\usepackage{listings}
\usepackage{wrapfig}
\usepackage{xspace}
\usepackage{enumitem}
\usepackage[compact]{titlesec}
\usepackage{mathpazo,euler}
\usepackage{multirow}
\usepackage{multicol}
\usepackage{tabularx}
\usepackage{array}
\usepackage[justification=centering]{caption}

\usepackage{amssymb}
\usepackage{ifthen}

\newboolean{showcomments}

\setboolean{showcomments}{true}

\ifthenelse{\boolean{showcomments}}
  {\newcommand{\nb}[2]{
    \fbox{\bfseries\sffamily\scriptsize#1}
    {\sf\small$\blacktriangleright$\textit{#2}$\blacktriangleleft$}
   }
   
  }
  {\newcommand{\nb}[2]{}
   
  }

\newcommand{\eg}{\textit{e.g.},\xspace}

\titlespacing{\section}{0pt}{2ex}{1ex}
\titlespacing{\subsection}{0pt}{1ex}{0ex}
\titlespacing{\subsubsection}{0pt}{0.5ex}{0ex}

\newcounter{taskctr}

\definecolor{dkgreen}{rgb}{0,0.6,0}
\definecolor{gray}{rgb}{0.5,0.5,0.5}
\definecolor{mauve}{rgb}{0.58,0,0.82}
\definecolor{shadecolor}{rgb}{0.95,0.95,0.95}
\makeatletter
\renewcommand{\paragraph}{%
	\@startsection{paragraph}{4}%
	{\z@}{0.5ex \@plus 0ex \@minus .2ex}{-1em}%
	{\normalfont\normalsize\bfseries}%
}
\makeatother

\makeatletter
\g@addto@macro\normalsize{%
	\setlength\abovedisplayskip{0pt}
	\setlength\belowdisplayskip{0pt}
	\setlength\abovedisplayshortskip{-10pt}
	\setlength\belowdisplayshortskip{0pt}
}
\makeatother

\DeclareGraphicsExtensions{.pdf,.png}
\graphicspath{{img/}}

\newcommand*{\img}[1]{%
    \raisebox{-.5\baselineskip}{%
        \includegraphics[
        height=1cm,
        width=1cm,
        keepaspectratio,
        ]{#1}%
    }%
}

\newcolumntype{C}[1]{>{\centering\arraybackslash}p{#1}}

\newcommand{\bfN}{\mathbf{N}}

\newcommand{\ignore}[1]{}

\renewcommand{\maketitle}{
\begin{center}

\pagestyle{empty}
\phantom{.}  
\vspace{2cm}

{\LARGE Deep Learning \& Software Engineering: State of Research and Future Directions}
\vspace{1.8cm}

{\large Prem Devanbu\footnotemark[3], Matthew Dwyer\footnotemark[4], Sebastian Elbaum\footnotemark[4], Michael Lowry\footnotemark[9], Kevin Moran\footnotemark[1], Denys Poshyvanyk\footnotemark[1], Baishakhi Ray\footnotemark[2], Rishabh Singh\footnotemark[8], and  Xiangyu Zhang\footnotemark[7]}
\vspace{2em}

{\footnotemark[3]University of California, Davis \hspace{0.5cm} \footnotemark[4]University of Virginia \hspace{0.5cm} \footnotemark[9]NASA Ames Research Center \hspace{0.5cm} \footnotemark[1]William \& Mary \hspace{0.5cm} \footnotemark[2]Columbia University   \hspace{0.5cm}  
\footnotemark[8]Google \hspace{0.5cm} \footnotemark[7]Purdue University}

\vspace{0.75em}

{Workshop participants: 
Christian Bird, Satish Chandra, Abram Hindle, Ranjit Jhala, Gail Kaiser, Bo Li, Zhenming Liu, Mike Lowry, Shiqing Ma, Collin McMillan, Tim Menzies, Audris Mockus, Raymond Mooney, Vijayaraghavan Murali, Nachi Nagappan, Koushik Sen, Charles Sutton, \\ Lin Tan, Danny Tarlow, Aditya Thakur, and Bogdan Vasilescu$^{+}$}
\vspace{2em}

{$^{+}$The affiliations of all the workshop participants are listed at \textit{https://dlse-workshop.gitlab.io/participants/}}

\vspace{3cm}
{\Large April 2020}

\vspace{4cm}
{\large Sponsored by the National Science Foundation (NSF), USA}

\newpage

\end{center}
}\makeatother

\begin{document}
\maketitle

\pagestyle{plain}

\suppressfloats
\vspace{1em}
\noindent{\Large \textbf{Executive Summary}}
\vspace{1em}

The advent of deep learning (DL) has fundamentally changed the landscape of modern software. Generally, a DL system is comprised of several interconnected computational units that form "layers" which perform mathematical transformations, according to sets of learnable parameters, on data passing through them. These architectures can be ``trained'' for specific tasks by updating the parameters according to a model's performance on a labeled set of training data. DL represents a fundamental shift in the manner by which machines learn patterns from data by automatically extracting salient features for a given computational task, as opposed to relying upon human intuition. These DL systems can be viewed as an inflection point for software development, as they enable new capabilities that cannot be realized cost-effectively through "traditional" software wherein the behavior of a program must be specified analytically. This has ushered in advancements in many complex tasks, often associated with artificial intelligence, such as image recognition, machine translation, language modeling, and recently, software engineering. DL is fundamentally intertwined with software engineering (SE), primarily according to two major themes.

 The first of these two themes is related to DL-techniques when viewed as a new form of software development, and in this report we refer to this as \textbf{Software Engineering for Deep Learning (SE4DL)}. In essence, the application of DL to a computational problem represents a new programming paradigm: \textit{rather than analytically specifying a program in code, a program is "learned" from large-scale datasets}. This new form of development carries with a new set of challenges that represent several opportunities for novel research.

The second of these two themes is related to leveraging Deep Learning techniques in order to automate or improve existing software development tasks, which we refer to as \textbf{Deep Learning for Software Engineering (DL4SE)}. There currently exists an unprecedented amount of software data that is freely available in open source software repositories. This data spans several different types of software artifacts, from source code and test code, to requirements and issue tracker data. Given the effectiveness by which DL systems are able to learn representations from such large-scale data corpora, there is ample opportunity to leverage DL techniques to help automate or improve a wide range of developer tasks. However, with this opportunity also comes a number of challenges, such as curating datasets to develop techniques for particular development tasks, and designing DL models that effectively capture the inherent structure present in a wide range of different software artifacts.

Given the current transformative potential of research that sits at the intersection of DL and SE, an NSF-sponsored community workshop was conducted in co-location with the 34th IEEE/ACM International Conference on Automated Software Engineering (ASE'19) in San Diego California. The goal of this workshop was to outline high priority areas for cross-cutting research that sits at the intersection of Deep Learning and Software Engineering. While a multitude of exciting directions for future work were identified, we provide a general summary of the research areas representing the areas of highest priority which are expanded upon in Section \ref{sec:introduction}. The remainder of the report expands upon these and other areas for future work with high potential payoff.

\vspace{1em}
\noindent \textbf{\large Community Research Challenges}

\begin{itemize}
    \item{\textit{\textbf{Laying the Foundations for SE4DL including defining DL workflows, proper DL abstractions, and theory for DL development.} - }  A systematic taxonomy and understanding of different development workflows for DL-based systems and how these differ from traditional software development practices is needed. Researchers should also look to develop salient DL program abstractions that can be easily constructed and analyzed in a formal manner.}
    \item{\textit{\textbf{Identifying failure modes of DL Systems and their corresponding countermeasures} - } There is a dire need for a detailed categorization and understanding of common types of faults for DL-based systems alongside automated techniques for detecting, debugging, and fixing such faults.}
    \item{\textit{\textbf{Furthering DL4SE through the development of tailored architectures, use of heterogeneous data sources, focusing on new SE tasks, and combining DL with existing empirical data.} - } New DL models designed specifically for given SE tasks that learn orthogonal information from a heterogeneous software artifacts while making use of the specific structural properties of such artifacts. Furthermore, researchers should look for ways to combine existing empirical knowledge into approaches for DL4SE, examine new categories of tasks, and look towards incorporating multi-modal data representations.}
    \item{\textit{\textbf{Establishing educational and pedagogical materials to better support training related to DL-based development via academic-industry partnerships} - } While progress is being made on understanding DL-based development workflows, there should be parallel effort dedicated to developing effective pedagogical and educational material to transfer newly discovered knowledge on to students. Industry-academic partnerships could aid in ensuring the immediate impact of such material.}
\end{itemize}

\vspace{1.5em}
\noindent \textbf{\Large \img{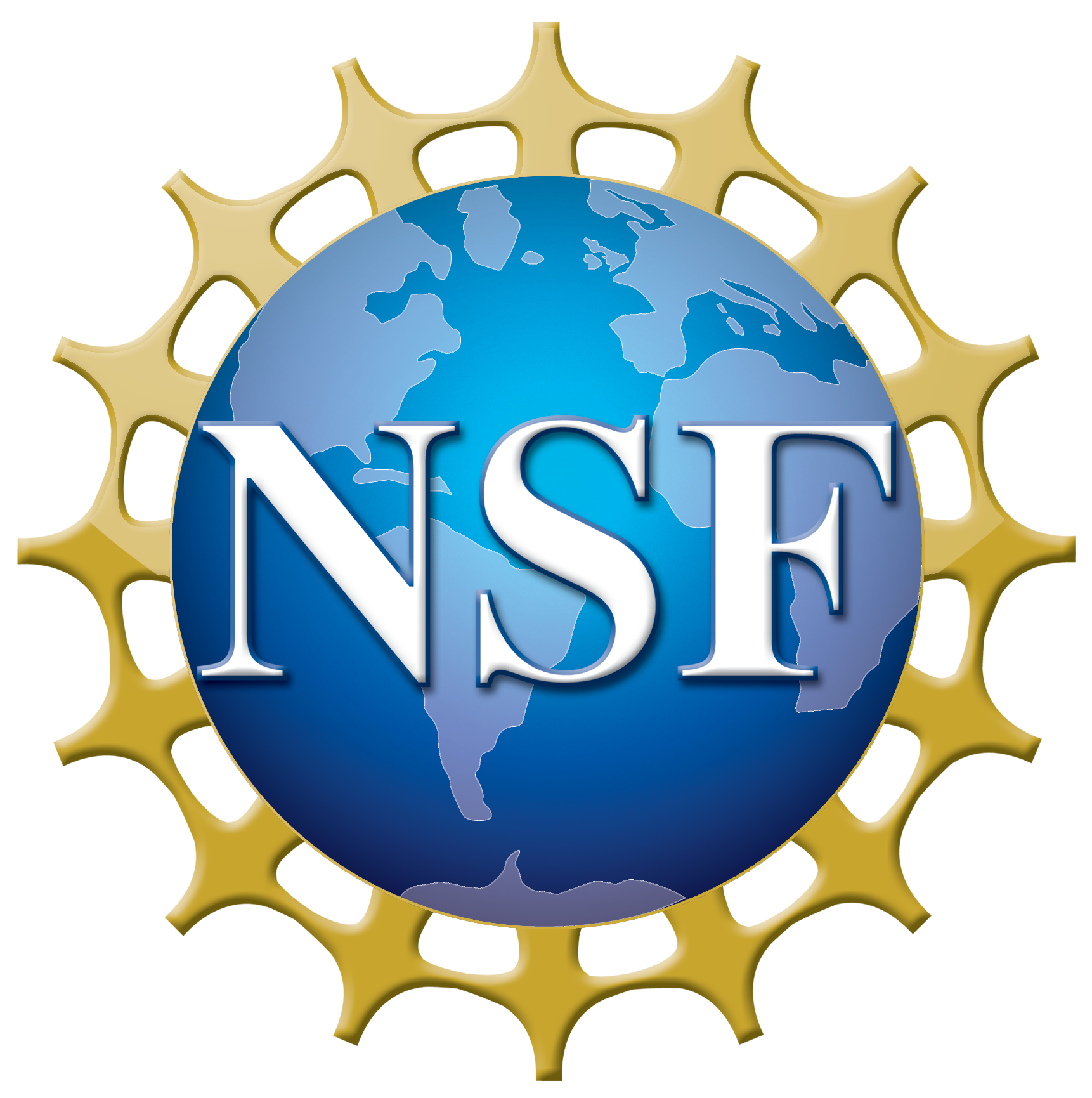} Acknowledgement of Sponsorship}
\vspace{1em}

This material is based upon work supported by the NSF under Grants No. CCF-1927679 and CCF-1945999. Any opinions, findings, conclusions, or recommendations expressed in this material are those of the authors and do not necessarily reflect the views of the NSF. We would also like to specifically acknowledge and thank Dr.~Sol Greenspan, sponsoring NSF program director, for support and guidance through the planning and execution of the workshop.

\newpage
\section{Introduction}
\label{sec:introduction}

\subsection{The Interplay between Deep Learning \& Software Engineering}
\vspace{0.5em}

While DL systems bring with them the potential to tackle problem domains generally considered to be outside the realm of "traditional" programs, they represent a radical shift in the manner in which software is created. These changes encompass the emergence of new development workflows, new development artifacts, and new program properties that are, as of yet, not well understood. For example, DL models exhibit (i) a high degree of coupling between different model components, (ii) inherent stochasticity, (iii) distinct computational encodings, (iv) extreme sparseness in their representations of data, and (v) specifications that are inherently difficult to pin down. Because of this, many of the software engineering (SE) techniques and processes that have been developed and refined over decades for traditional software systems are not directly applicable to DL systems. For instance, techniques assuming deterministic behavior of software might fail due to the stochasticity of DL models whereas techniques that might assume a program is well-defined over most of its input domain (such as fuzzing) might fail due to the sparseness. Thus, there are clear challenges that exist related to understanding and supporting a new software development paradigm for DL systems, which we will refer to as Software Engineering for Deep Learning (SE4DL).

While challenges exist in adapting SE techniques and practices to DL systems, the capabilities that DL systems also  offer a tremendous opportunity to improve or automate several aspects of the "traditional" software development process. While DL provides a powerful solution to certain complex computational intelligence tasks, it is likely that "traditional" analytical programming methods will be more cost-effective for easily specifiable tasks, such as interfacing with a database. Thus, it would be preferable to automate or improve developer's effectiveness in these tasks. DL is poised to offer transformative techniques for traditional software development due to (i) the scale of software artifact data (e.g., code) in online repositories, (ii) the automated feature engineering provided by DL techniques, (iii) the robustness and scalability of optimization techniques such as gradient descent, and (iv) the transfer-ability of traditional DL applications (such as language modeling) to SE artifacts. These factors indicate the great potential for DL to tangibly improve the traditional software development process, and we refer to this area of research as Deep Learning for Software Engineering (DL4SE).

\subsection{Summary of Research Opportunities in SE4DL}
\vspace{0.5em}

In this report, we identify pertinent challenges related to \textbf{SE4DL} that were discussed at the workshop and motivate fundamental research to:

\begin{itemize}

	\item{\textbf{Expose the structure, variability, and characteristics of existing workflows for creating DL systems.} Decades of empirical research on the development workflows for traditional software engineering systems have provided a wealth of knowledge for such processes. However, given the nascent state of DL-based software systems, there is a need for additional empirical work with the aim of understanding emerging DL-based development workflows. Such work will require close cooperation between academic and industrial researchers, as indicated by early studies~\cite{Amershi:ICSE'19}.}
	
	\item{\textbf{Identify the abstractions inherent in DL workflows, models, and implementations which form the basis for SE of DL.} Much of the automated support for traditional software systems is driven by mature research and techniques for \textit{program analysis}.  These program analysis techniques are often built upon analyzing and manipulating abstract representations of traditional software programs (e.g. using abstract syntax trees or control-flow graphs). Thus, before work can begin in earnest on new classes of program analysis techniques for DL-based systems, researchers need to understand and define appropriate abstractions that are amenable to more advanced analysis and manipulation than model code or weights.}
	
	\item{\textbf{Identify modes of failure and their counter-measures within these workflows.} While failures and faults in traditional software systems tend to be quite varied, empirical research on understanding software bugs and testing has provided a rich knowledge base of faults across different types of systems. This has led to the formulation of automated techniques for identifying, reporting, triaging, and fixing such faults. However, the failure modes of DL-based systems currently are not well understood. For instance, faults may be highly coupled to the task a given model is employed for, and it may not be clear whether a given fault maps back to problems with a given model or the dataset the model learned from. Further research is needed to better understand the nature of faults for DL-based systems so that new techniques for testing and verification can be developed.}
	
	\item{\textbf{Establish theoretical foundations upon which cost-effective SE techniques for DL systems can be built.} Formal program analysis techniques have become a key building block for scalable and accurate analysis of software systems. However, proper theoretical underpinnings for DL-based software are needed in order to help help drive the next generation of these techniques. For example, defining implicit model-level property specifications is a necessary next step towards driving advancements in the verification and validation of DL-based systems.}
	
\end{itemize}

\subsection{Summary of Research Opportunities in DL4SE}
\vspace{0.5em}

Additionally, we identify several promising directions for future work related to \textbf{DL4SE} including:
\begin{itemize}
	\item{\textbf{Combining features learned from large-scale SE data with empirical human knowledge to more effectively solve SE tasks.} Decades of research on traditional SE processes and techniques has led to the development of a sizeable knowledge base regarding best practices for various tasks, attributes of effective tools, and understanding of human-centered processes. While DL techniques have shown immense potential in automatically learning features from large datasets, the empirical knowledge from the SE community should not be \textit{completely} ignored, as such synthesized knowledge may not necessarily be captured by DL techniques. Thus, there is need to develop creative techniques for combining the learned features of DL-based models with empirical knowledge to build more effective automated approaches. Such research could manifest the in the creation of labeled training datasets drawn from existing knowledge bases and taxonomies or in guiding DL techniques to learn specific features through targeted data pre-processing.}
	
	\item{\textbf{Leveraging heterogeneous sources of SE data (source code, requirements, issues, visual artifacts).} While code is generally regarded as the essence of a software system, in no way is it the only representation of software that developers handle in their daily workflows. In fact, software data is contained across three major information modalities: (i) \textit{code} represents source code and its corresponding abstractions;} (ii) \textit{natural language} manifests across a variety of software artifacts such as requirements, comments, issue trackers, bug reports, among others; finally (iii) \textit{graphical} artifacts are also abundant in software, comprising user interfaces and design documents among others. Given the inherent diversity among these representations, it stands to reason that these different information modalities capture \textit{orthogonal} properties or knowledge about a given underlying software system. Thus, future work should look for creative methods of combining these information sources together for richer DL-based representations of software.
	
	\item{\textbf{Developing tailored architectures that exploit the unique properties of SE data in order to offer better automated support to developers.} DL techniques have largely focused upon sequence-based learning for natural language corpora and spatial-based learning for graphical data. While such techniques can be applied "out of the box" for SE data such as code, requirements, or graphical data, artifacts of traditional software systems exhibit generally well-understood structural properties (e.g., control-flow in code and layouts of widgets in graphical user interfaces). As such, there is a need for researchers to develop techniques that take advantage of these structural properties in order to learn more effective DL-based representations of software artifacts for a variety of uses.}
	
	\item{\textbf{Defining a systematic and reproducible research methodology for DL applied to traditional SE tasks.} Reproducible and replicable research is the driving factor of scientific discovery. While all scientific disciplines face challenges related to the trade-offs of new discoveries versus the verifiability of past results, research involving DL techniques poses further challenges to reproducibility. Extremely large scale datasets, stochastic training processes, and variability in data preprocessing all contribute to unique difficulties in reproducibility of research related to DL4SE. Thus, rigorous research methodologies and transparency in the research process are essential.}

\end{itemize}

\subsection{Summary of Cross-Cutting Research Opportunities}
\vspace{0.5em}

Finally, this report outlines workshop discussions of several concerns, serving as topics for future work, that cross-cut both SE4DL and DL4SE:

\begin{itemize}

	\item{\textbf{Developing methods to explain how DL-based systems arrive at predictions.} Given their complex, high dimensional representations of data, it can be difficult to interpret a given prediction made by a modern DL-based software system. Such opaqueness complicates research that cross cuts both SE4DL and DL4SE. For instance, understanding common failure modes of DL systems will be difficult without some degree of model interpretability, furthermore, DL-based tools to automate traditional development tasks may be difficult to interpret in practice. Thus, interdisciplinary research on model explainability is needed.}

    \item{\textbf{Education for both students who will seek emerging positions that require engineering systems enabled by both DL and software technologies, and researchers who will study the inter-play of DL and SE.} Given the rapidly growing popularity of DL-based software systems, it is imperative that educational materials for both students and researchers are developed and widely disseminated. For students, coursework on both fundamental machine learning principles, as well as effective software engineering practices for DL-based systems will be important areas of curriculum development. For researchers, materials that offer guidance on rigorous empirical methods for carrying out research at the intersection of DL and SE are needed.}

    \item{\textbf{Community infrastructure for research in DL applied to SE, and for investigating the engineering fundamentals of DL systems.} A shared community infrastructure for managing a variety of DL-based artifacts such as models, code, and evaluation metrics would drastically improve the reproducibility of cross-cutting research.} 

    \item{\textbf{The need to foster a cohesive research community, involving both academics and industrial practitioners, in order to move the field forward.} The need for academic-industrial partnerships to advance several of the research directions above is immediately clear. For work on SE4DL, there is a need to study and understand current industrial practices and applications, and prove out new methodologies or tools in practice. For work on DL4SE, industrial datasets can be invaluable, as can access to evaluate DL-driven automation with real developers. Such partnerships should be a major goal for research moving forward.} 

\end{itemize}

\subsection{Report Structure}
\vspace{0.5em}

    The remainder of this report is structured in a manner that mirrors the breakout sessions of the 2019 NSF Workshop on Deep Learning and Software Engineering (Section~\ref{sec:workshop}). That is, each section following the introduction is dedicated to summarizing the discussion and crystallizing detailed directions for future work as specified by the attendees of the workshop. These directions are delineated in the text as discrete points labeled according to the overarching research topic in which they are situated. These sections largely expand upon the summary of research opportunities that are discussed above and in the \textit{Executive Summary}.
\section{Deep Learning for Software Engineering}
\label{sec:dl4se}

\subsection{Background}
\vspace{1em}

Software engineering (SE) research investigates questions pertaining to the design, development, maintenance, testing, and evolution of software systems. As software applications pervade a wide range of industries, both open- and closed-source code repositories have grown to become unprecedentedly large and complex. This results in an increase of unstructured, unlabeled, yet important data including requirements, design documents, source code files, test cases and defect reports. Previously, the software engineering community has applied traditional machine learning (ML) techniques to identify interesting patterns and unique relationships within this data to automate or enhance many tasks typically performed by developers. Unfortunately, the process of implementing ML techniques is often a tedious exercise in careful feature engineering, wherein researchers experiment with identifying salient pieces of data that can be leveraged to help solve a given problem or automate a given task. 

Due to recent improvements in computational power and the amount of memory available in modern computer architectures, the rise of DL has ushered in a new class of learning algorithms particularly suited for large datasets. DL represents a fundamental shift in the manner by which machines learn patterns from data by \textit{automatically} extracting salient features for a given computational task, as opposed to relying upon human intuition. Deep Learning approaches are characterized by ``architectures'' comprised of several ``layers'' that perform mathematical transformations, according to sets of learnable parameters. These computational layers and parameters form models that can be trained for specific tasks, such as image classification, by updating the parameters according to a model's performance on a set of training data. Given the immense amount of data in software repositories that can serve as training data, deep learning techniques have ushered in advancements across a range of tasks in software engineering research including automatic software repair~\cite{Tufano:2018:EIL:3238147.3240732}, code suggestion~\cite{Gu:2018:DCS:3180155.3180167}, defect prediction~\cite{7886912}, malware detection \cite{8433204}, feature location~\cite{7332513} among many others~\cite{Ma:2018:DMT:3238147.3238202, Wan:2018:IAS:3238147.3238206, Liu:2018:DLB:3238147.3238166, White:2016:DLC:2970276.2970326, Xu:2016:PSL:2970276.2970357, 7985645, 8453089, 7985701}. This field of research, which we refer to as DL4SE, shows clear potential for transforming the manner by which a variety of specific "traditional" software development tasks are performed. However, there are still several open questions that represent promising opportunities for research moving forward.

\subsection{Research Opportunities}
\vspace{0.5em}

\subsubsection{Identifying Applicable SE Tasks}
\vspace{0.5em}

One of the first questions that a researcher must ask when working on DL4SE topics is the SE task to which a DL approach will be applied to improve, automate, or study. Recent work from the SE community~\cite{Fu:FSE'17} seems to suggest that applying DL to automate certain SE-related tasks may not provide improvements in either effectiveness or efficiency of techniques. Therefore, researchers must critically analyze the problem domain that they plan to target in order to decide whether or not a DL-based solution is appropriate given the context, and should always compare against a computationally "simpler" baseline. For instance, it may be better to apply more formal static analysis techniques to provide certain guarantees of program behavior, as opposed to DL techniques. Generally, DL models tend to be better applied in tasks where the relevant features for delineating patterns are difficult to engineer, or pin down analytically. This being said, there are a number of un- or under-explored SE tasks that may serve as promising future applications.

\begin{quote}
\textbf{DL4SE$_1$:}\textit{ Potential SE tasks identified as showing promise for automation or improvement via DL techniques include: (i) software testing (in various forms), (ii) troubleshooting tasks (e.g. incident management, resolving deployment bugs), (iii) bug triaging, and (iv) code review. More broadly, any SE task for which it is possible to collect data, but difficult to engineer salient features may be ripe for DL applications.}
\end{quote}

While choosing an appropriate SE-related problem domain is important to ensure that a DL application is appropriate, there are also downstream questions that must be answered regarding the usability of such systems and fit into modern development workflows. Ultimately, any automated system that is designed with DL underpinnings must somehow fit into the day-to-day workflow of developers and fundamentally improve the specified task. As such, researchers should be consciously thinking about how the tool they are building will fit into such workflows, and ultimately interface with developers or other relevant software practitioners. The general experience from industrial practitioners thus far has been that tools or approaches that are designed to work \textit{synergistically} with developers tend to be the ones that are most successful in practice. Automatically performing tasks outright without the potential for developers to intervene can lead to distrust and slow or low adoption. Thus, there is clear need both for existing research on DL-powered tools to consider its practical applicability, and for researchers to study and better understand how to enable synergistic relationships between these DL-powered tools and developers \textit{within the context of specific SE tasks}.

\begin{quote}
\textbf{DL4SE$_2$:}\textit{ Research into understanding and enabling synergistic relationships between developers and DL-powered SE tools and techniques is needed in order to ensure that such tools are able to effectively support developers in practice.}
\end{quote}

\subsubsection{Data Resources in DL4SE Research}
\vspace{0.5em}

DL techniques are inherently data-centric. As such, the various types of SE data that are being used to drive advancements in DL4SE research are vital to research workflows. As software has continued to grow in its pervasiveness, researchers have been met with a marked increase in the availability of open source software data freely hosted in online repository ecosystems such as GitHub, GitLab, and Sourceforge, just to name a few. The data contained within these software repositories is not solely comprised of code, instead they typically contain a variety of artifacts such as requirements, issues, design documents and even graphical data such as screenshots and videos of user interfaces. While these artifacts represent a rich set of resources to which SE researchers can apply DL, there are a variety of challenges in properly "cleaning" or preparing the data such that relevant patterns can be learned for a particular SE task. The process of sourcing large-scale datasets from open source software repositories brings with it potential pitfalls (e.g., the dangers of code duplication~\cite{Allamanis:SPLASH19}) that must be identified and studied in order to better understand their effects on downstream SE tasks.

\begin{quote}
\textbf{DL4SE$_3$:}\textit{ There is a need for further research that examines the methods by which large scale datasets are curated from open source software repositories and "cleaned", including the potential (negative or positive) implications on DL-based techniques aimed at improving or automating downstream SE tasks.}
\end{quote}

Perhaps the most popular form of DL techniques that are applied in practice today are those based on the concept of \textit{supervised learning}, wherein patterns are learned from data that has been "labeled" with explicit patterns to be learned by the DL system (e.g., labeling images with categories that describe the image). Curating large labeled datasets is typically a time and effort-intensive undertaking, as it requires \textit{manual} labeling of discrete pieces of data. However, due to the structure inherent in software artifacts there is the potential for synthesizing \textit{automatically} labeled datasets by leveraging this structure. For example, compilers could be used as an "oracle" of sorts for automatically labeling code that is structurally sound, and different types of code constructs could be automatically labeled through representations as Abstract Syntax Trees (ASTs).

\begin{quote}
\textbf{DL4SE$_4$:}\textit{ There are ample research opportunities for synthesizing automatically labeled datasets for SE tasks by exploiting the inherent structure present in a range of software artifacts.}
\end{quote}

The variety of software artifacts that are available in open source repositories represent an interesting research opportunity for SE researchers. Broadly speaking, the various artifacts that exist (source code, test code, issue tracking data, screenshots, logs, requirements, etc.) belong to one of (or a mixture of) three different information modalities: (i) code, (ii) natural language, and (iii) graphical information. Software systems are complex, and generally leverage abstractions to help developers mentally manage the complexity of a given system. Given the diversity in different software artifacts, and their corresponding representations in the information modalities listed above, it stands to reason that these different representations capture orthogonal attributes or aspects across different software abstraction levels. As such, this would signal the need for research that both attempts to understand \textit{what} aspects of software these different artifacts capture, and \textit{how} different artifacts across information modalities might be combined together in order to form more descriptive datasets for DL-based techniques. Past work has shown the potential of combining multiple information sources for tasks related to machine understanding~\cite{Moran:TSE'18,LeClair:ICSE'19}, and if properly applied to diverse collections of SE data, promising new progress could be made. 

\begin{quote}
\textbf{DL4SE$_5$:}\textit{ Future DL4SE work should look for opportunities to combine information from different software artifacts and modalities of information in order to build more descriptive datasets to support the development of tailored DL techniques applied to SE tasks.}
\end{quote}

While the diverse array of open source data provides a promising vehicle by which DL4SE research can be driven, there is a limit to the types of data which can be collected from them. For instance, it would be difficult to mine fine-grained developer interactions from modern open source software repositories. With this in mind, there is great potential for mining information from new software related sources. One particularly promising information source may be often-used developer tools, and in particular the IDE. Collecting fine-grained information from IDEs is not an strictly new direction of research~\cite{Robbes:ICPC07}, however its application in synthesizing datasets for DL-based tools represents a promising avenue of work. For instance, one could envision a proactive IDE assistant that is able to predict context switches between common development tasks and provide developers with optimized suggestions based upon a chain of past fine-grained interactions.

\begin{quote}
\textbf{DL4SE$_6$:}\textit{ Fine-grained instrumentation of often used development tools, such as IDEs, could provide a new, richly detailed source of data to drive DL4SE research. Therefore, future work should examine proper methods of instrumentation to capture relevant data in a privacy-sensitive manner, and look to utilize this data to drive new innovations for DL-assisted developer tools.}
\end{quote}

\subsubsection{Software Artifact Specific DL Models \& Architectures}
\vspace{0.5em}

As stated above, software artifacts, and in particular code, carry with them inherent structural properties that tend to make them unique compared to other forms of unstructured data. However, until very recently, much of the work in applying DL to these structured artifacts has been conducted by applying models primarily targeted at data types that are inherently less structured (e.g., sequence-based language models for natural language). Thus, there is a clear opportunity for future work to design new types of DL architectures that represent or encode software-related artifacts in such a manner that take advantage of their inherent structure or properties. Recent applications of graph- and tree- based neural networks applied to source code represent an early step in this direction. Additionally, the properties of many existing DL architectures may not properly align with target SE tasks. For example, many current models for machine translation are specifically engineered to be "meaning preserving" between translated phrases. However, this property could prove troublesome for certain SE tasks, such as defect repair, given that the translated phrases must by definition have different meanings. Due to such properties of existing DL architectures, future work should also look into their viability for SE data and the potential advancements made by SE-specific architectures. 

\begin{quote}
\textbf{DL4SE$_7$:}\textit{ Given the unique context and structure of the many different types of software artifacts, future work should focus upon understanding the implications of applying DL architectures from other domains on SE data and designing architectures that are capable of representing SE-related artifacts in a manner that leverages their unique properties for more effective learning.}
\end{quote}

As with any type of dataset, in datasets involving SE-related data there will be a distribution of different labeled examples that will impact what is learned by the corresponding models. For example, if one were to task a DL-based system to learn to fix bugs based on a dataset mined from GitHub, it would be expected that such a DL-based system would learn to identify and fix \textit{common} bugs in a relatively effective manner, whereas lesser seen bugs may give the system more trouble. In the context of automating different software engineering tasks, this phenomenon of the \textit{long tail} of potentially rare examples should be considered in relation to its practical implications for the approach being developed. For instance, an automated bug fixing approach that is able to pinpoint problems that are easily identifiable by developers may be less valuable than one that is able to to uncover and fix bugs that are harder to spot, or that are of a higher severity level.

\begin{quote}
\textbf{DL4SE$_8$:}\textit{ Researchers should consider the effect of the "long-tail" of the predictive power of DL models when applied to data mined at scale with unknown distributions. Furthermore, future work should examine techniques for mitigating or understanding these effects when DL techniques are applied to SE tasks.}
\end{quote}

While there is clear potential benefit to customizing DL-based models to better fit SE-related data, there is also an open question of \textit{how such models should be evaluated}. Traditionally, work on DL4SE has used or adapted effectiveness metrics from the machine learning (ML) and natural language processing (NLP) research communities. However, it is not clear the extent to which this is appropriate for SE-related data and tasks. For example, adapting the relatively popular BLEU~\cite{bleu} metric to source code may not be the best method for measuring the quality of generated code due to its "naturalness" (i.e., repetitiveness) relative to other types of textual data. From this point of view, there is clear need for developing effectiveness metrics specifically for software-related data. For instance, this could be done by comparing a variety of existing and newly proposed metrics to human evaluators in order to determine the metric that best reflects effectiveness of a given task as perceived by a developer.

\begin{quote}
\textbf{DL4SE$_9$:}\textit{There is a need for research into designing proper effectiveness metrics for DL techniques that are specifically applied to learning patterns from software-related data, as current metrics from the ML and NLP communities may not be a good fit.}
\end{quote}

\section{Verification \& Validation of Deep Learning Systems}
\label{sec:verification}

\subsection{Background}
\vspace{0.5em}

Formal verification and validation approaches have become a key building block for scalable and accurate
analysis of software systems, but this did not happen over night.
For example, the first applications of symbolic execution to the validation of software
appeared in the mid 1970s, e.g., \cite{king1976symbolic,clarke1976system}, but it took three decades for the underlying
technologies, e.g., \cite{de:2008:z3,barrett-tinelli:2007:cvc3}, 
to mature to the point where researchers could push symbolic execution further
to apply it to realistic software systems or their components, e.g., \cite{khurshid-pasareanu-visser:2003:tacas,sen-etal:2005:cute,cadar2008klee}.
Today, these technologies are applied as a regular component of development workflows to validate software
and help verify the absence of faults and security vulnerabilities~\cite{DBLP:journals/cacm/CadarS13}.
We believe that developing cost-effective techniques for verifying and validating realistic deep
learning system models will require a sustained long-range program of research, and
we outline key elements of such a research program below.

\vspace{0.5em}
\subsection{Research Opportunities}
\vspace{0.5em}

\subsubsection{Property Specification}
\vspace{0.5em}

Verification and validation requires an explicit \textit{property specification} -- a precise
formal statement of the expected system behavior.
Specifying a property of a feed-forward DL model, $\bfN : \mathbb{R}^n \rightarrow \mathbb{R}^m$, involves the definition, $\phi \subseteq \mathbb{R}^n \times \mathbb{R}^m$, of the allowable inputs and their associated outputs.    
Since the process of specification is costly it is commonplace for verification and validation to 
exploit \textit{partial specifications}, which define necessary conditions for a system
to be correct.

Common sources of such specifications include the semantics of the programming language, e.g.,
freedom from null pointer dereference or out of bounds array indexing, or the
programming environment, e.g., reading is only permitted on a file that was
previously opened, in which the system is implemented.  Such \textit{implicit property specifications} are particularly valuable
because they are system-independent, for example, any system that dereferences
a null pointer is faulty.  This allows for such property specifications to be
formulated a single time and applied to verifying and validating any system, e.g., ~\cite{das2002esp,ball2004slam}.

DL models can suffer from faults related to violation of implicit specifications
in their implementations.   For example, the TensorFlow implementation of a model
might compute a NaN value when running inference on a particular input.  
Unlike existing work on
implicit specification, in this case such a violation may implicate
either the model or its implementation.   A key advance in validation and verification of DL models
will be to develop the analog of the ubiquitous ``freedom from null pointer dereference'' property for traditional 
software for DL models.

\begin{quote}
\textbf{VV$_1$:}\textit{ Research on defining implicit model-level property specifications for DL systems is needed to drive advances in and broad application of DL verification and validation approaches.}
\end{quote}

There is a rich literature on languages for stating \textit{explicit property
specifications}, e.g., assertions\cite{rosenblum1995practical}, 
contracts\cite{meyer1992applying,leavens1999jml}.
Such specifications are system-dependent and, consequently, they must be written
by system developers prior to verification and validation.

It is challenging to write specifications for traditional software, but the nature
of DL models introduces significant new challenges.   First and foremost, DL models
are  used when specifying the target function that the model aims to accurately
approximate is effectively impossible.   For example, specifying a Boolean classifier
for detecting the presence of a pedestrian in an image must account for 
the distance, orientation, occlusion, and natural variation of the shape and pose of a human --
not to mention myriad other factors.  Such models are thought to 
be \textit{inherently unspecifiable} which is surely true if one seeks a complete
specification of a DL model.
Yet researchers have made some progress by focusing on partial specifications, for example,
defining the maximum allowable steering angle for a regression network \cite{bojarski-etal:corr:2016:DAVE2}.

\begin{quote}
\textbf{VV$_2$:}\textit{ Research on defining explicit property specifications for DL systems should focus on partial specification of necessary conditions for correctness.}
\end{quote}

Despite the above challenges there has been progress in specifying properties of DL models.
Researchers have observed for some time that the continuity of a function means 
that \textit{metamorphic relations}, which state that small
changes in the input result in small changes in the output, should hold over
its input domain~\cite{liu2013effectively,murphy2008properties}.
The ML community has formulated a variety of such properties which 
are oriented towards determining a model's 
\textit{robustness to adversarial examples}~\cite{szegedy2013intriguing,goodfellow2014explaining,papernot2016practical}.   
Such specifications are sensible for regression models or when interpreting the values in the output layer of a categorical network.
However, when applied to the entire input domain robustness properties do not make sense for categorical models.
Such a property, e.g., $\forall i \in \mathbb{R}^n : \forall r \in [-\epsilon,\epsilon]^n : \bfN(i) = \bfN(i+r)$, implies by transitivity that the model can produce a single output value -- since the entire input domain can 
be spanned by overlapping regions of diameter $2 * \epsilon$.
This precludes classifiers that define a decision boundary and, consequently, the 
robustness is typically verified or validated on a very small sample of inputs relative to $\mathbb{R}^n$.

\begin{quote}
\textbf{VV$_3$:}\textit{ Research on understanding the value of metamorphic properties for categorical
networks that go beyond sampling would be valuable.}
\end{quote}

To address the challenge of writing property specifications, Ernst and colleagues developed
the concept of \textit{property specification inference} by observing system behavior and
generalizing it to a form that can be expressed as an assertion or contract.  Research
from the ML community on rule extraction, e.g., \cite{setiono1995understanding,bastani2018verifiable} to promote 
interpretability of models and explainability of their inferences shares a similar aim -- explicating
the complex inner workings of the system in a set of simpler partial specifications.
A first step towards property inference for DL models~\cite{gopinathASE19} has been made, however, this direction of work is still in its early stages.

\begin{quote}
\textbf{VV$_4$:}\textit{ Research on inferring concise specifications from model behavior that capture properties 
of models is a promising direction for overcoming the challenge of
specifying their behavior ahead of time.}
\end{quote}

\subsubsection{Adapting Validation to Deep Models}
\vspace{0.5em}

In the past two decades, a key advance in validating software systems has been the development of techniques
for forcing the execution or simulation of system behavior.  Whether
these techniques operate systematically, e.g., symbolic \cite{klee:osdi:2008} or concolic \cite{sen-etal:2005:cute} execution,
use randomized approaches, e.g., fuzzing \cite{afl}, their power lies in being fully automatic.
Algorithms and machines force the system through millions, or billions, of behaviors without developer
intervention.  In addition to reducing developer effort, this also eliminates bias that might cause
a developer to miss exercising a particular behavior and lead to latent system faults.   
Their ability to generate large numbers of behavior means, however, that is infeasible for developers
to determine whether the system output is correct -- there are simply too many behaviors to consider.
Consequently, these techniques rely on the availability of property specifications that can be encoded
into monitors that evaluate internal system states or externally visible system behavior relative to properties.

Researchers have begun the process of adapting these methods to DL models, e.g.,
\cite{sun-etal:ASE:2018,xie2019coverage,DBLP:conf/icml/OdenaOAG19,xie2019deephunter,DBLP:conf/icse/GopinathPWZK19}, but they would be much
more effective with a broad array of meaningful property specifications.
As a corollary to the research directions listed above

\begin{quote}
\textbf{VV$_5$:} \textit{ Research on automated test generation for deep models must evolve with and adapt to developments in property specification in order to maximize their impact in validating DL system behavior}
\end{quote}

\subsubsection{Scaling Verification to Realistic Models}
\vspace{0.5em}

It has taken nearly four decades to develop the foundations, algorithms, and efficient implementations for
verification and validation of realistic software systems \cite{DBLP:journals/cacm/CadarS13}.
The importance of DL models has led researchers to seek to adapt such approaches in recent years leading
to more than 20 different published verification techniques,
e.g., \cite{katz-etal:CAV:2017,DBLP:conf/atva/Ehlers17,8318388,ai2,DBLP:conf/ijcai/RuanHK18,NIPS2018_8278,tjeng2018evaluating,Bastani:2016:MNN:3157382.3157391,Dvijotham18,DBLP:conf/icml/WongK18,DBLP:conf/iclr/RaghunathanSL18,DBLP:conf/nips/WangPWYJ18,DBLP:conf/uss/WangPWYJ18,DBLP:conf/icml/WengZCSHDBD18,huang:2017:CAV,Boopathy2019cnncert,DBLP:conf/nfm/DuttaJST18,DBLP:conf/nips/BunelTTKM18},
spanning three major algorithmic categories~\cite{dnnverificationsurvey}.
The pace of innovation in DNN verification is promising, but to date these techniques cannot scale
to realistic DL models -- they either exceed reasonable time bounds or produce inconclusive results.
Consequently, in applying the techniques developers
restrict property specifications to very small fragments of the
input domain to gain a measure of tractability \cite{DBLP:conf/nips/WangPWYJ18,NIPS2018_8278,Dvijotham18},
restrict input dimension to facilitate verification  \cite{DBLP:conf/nips/WangPWYJ18,DBLP:conf/ijcai/RuanHK18}, and only consider networks
with a modest number of layers and neurons that do not reflect the rapidly increasing DNN complexity \cite{katz-etal:CAV:2017,DBLP:conf/nips/BunelTTKM18,katz2019marabou}.

Scaling verification and validation  for traditional software was achieved in large part through
the application of frameworks for \textit{abstracting system behavior} and performing \textit{compositional
reasoning} that divides the system into parts, reasons about them, and combines their results to 
reflect system behavior.

Several DNN verification approaches have explored the use of abstraction, e.g., \cite{singh-etal:POPL:2019:deeppoly,DBLP:conf/nips/WangPWYJ18}, to 
soundly over-approximate model behavior and thereby permit efficient verification.  As with traditional
software, the key to abstraction is to control the over-approximation so as to preserve the ability to
prove properties -- coarse over-approximation leads to inconclusive \textit{unknown} results in verification.
One framework for achieving this in traditional systems is \textit{counter-example guided abstraction
refinement} \cite{clarke2000counterexample} which systematically and incrementally customizes the abstraction based upon both
the property and the structure of the system being verified.

\begin{quote}
\textbf{VV$_6$:}\textit{ Research into abstraction refinement approaches for the verification and validation 
of DL models is needed to scale to realistic models while preserving the accuracy of verification results.}
\end{quote}

Reasoning about large software systems requires the ability to divide and conquer.  In data flow analysis,
this is achieved through sophisticated frameworks for inter-procedural analysis that summarize the behavior
of individual functions and then incorporate them into system level reasoning with just the right measure of context
to allow for accurate results \cite{nielson-etal:2005:principles-of-pa}.   
The research community has realized that ``whole program'' verification and validation 
of software is impractical and there is no reason to believe that  ``whole DL model'' verification
and validation will fare any better.
However, DL models are built out of components -- a graph of layers of varying type.
While the nature of these components and their interface to one another varies in significant
ways from traditional systems, the component-based nature of DL models may be ripe for exploitation.

\begin{quote}
\textbf{VV$_7$:}\textit{ Research into compositional verification and validation of DL models is needed to
scale to realistic DL models.}
\end{quote}

\subsubsection{Benchmarks and Evaluation}
\vspace{0.5em}

Advances in verification and validation techniques for traditional systems have
resulted from an ongoing interplay between theoretical and empirical work.
An analysis of progress in SAT solving over a period of 6 years convincingly demonstrates
how regular empirical evaluation of the state-of-the-art drives the consolidation of the
best ideas \cite{barrett20136}.
Benchmarks are a necessary ingredient in enabling such evaluation and other verification communities, e.g.,
for SMT \cite{barrett2010smt} and theorem proving \cite{tp-bench}, have long recognized this and invested in their development.

Most papers published on DL model verification and validation use only a small set of \textit{verification
problems} -- a pair of a model and a property specification.  For example, the ACAS network from the
landmark paper by Katz et al.~\cite{katz-etal:CAV:2017} is still used, e.g, \cite{katz2019marabou,gopinathASE19}, despite the fact
that it has orders of magnitude fewer neurons than realistic models, e.g., \cite{loquercio-etal:RAL:2018:dronet,bojarski-etal:corr:2016:DAVE2} -- 
not to mention that it only includes fully connected layers.   

\begin{quote}
\textbf{VV$_8$:}\textit{ Research on developing corpora of verification problems that represent important classes
of realistic DL models are needed in order to evaluate techniques and drive algorithmic and implementation 
improvements.}
\end{quote}

The existence of benchmarks is not enough.  The research community must agree to use them.
Appropriate incentives must be put in place to encourage this, for example, requiring that any work accepted
for publication perform direct comparison with alternate approaches on benchmarks.  In other verification
fields establishing yearly competitions has been successful in building such community 
expectations and in highlighting the state-of-the-art \cite{hw-comp,sat-comp,smt-comp,sv-comp}.

\begin{quote}
\textbf{VV$_9$:} \textit{ Researchers should consider establishing regular competitions for DL model verification and validation techniques.}
\end{quote}

\subsubsection{Design for Verification}
\vspace{0.5em}

It is well-understood that verification and validation approaches are undecidable in general, but
researchers have pursued them for traditional software because they need only be cost-effective for the software
that people write.   Taking this line of reasoning further, researchers in high-confidence systems have
placed restrictions on the structure of software systems that lend the results amenable to automated
verification and validation \cite{chapman2014we,holzmann2018power,crocker2007verification}.   To date DL research has focused primarily
on improving the test accuracy of models and, to a lesser extent, their robustness to adversarial
examples.  As DL model verification and validation matures and an understanding
of what types of model structures simplify and complicate verification there is an opportunity
to bias model architecture to facilitate verification.

\begin{quote}
\textbf{VV$_{10}$:} \textit{ Research exploring how model architecture facilitates or complicates verification and validation could pave the way for developers to design models for verification.}
\end{quote}

\subsubsection{Beyond Feed-forward Models}
\vspace{0.5em}

The research on DL model verification and validation described above has focused on feed-forward models, but
there are other DL paradigms, such as DRL and RNN, that have received some attention, but deserve much more.
For example, recent work has applied the concept of \textit{policy extraction} to DRL models to extract
an alternative model, e.g., a program fragment \cite{DBLP:conf/icml/VermaMSKC18}, a decision tree \cite{bastani2018verifiable}, that is
amenable to verification using existing techniques.

The sequential nature of these paradigms presents challenges for verification and validation, but these
are not unfamiliar challenges.   Distributed and concurrent programs exhibit the same characteristics
and research on their verification and validation gave rise to temporal logics for specification~\cite{mannaTLbook} and 
model checking~\cite{clarke-grumberg-peled:1999:book} for verification.

\begin{quote}
\textbf{VV$_{11}$:} \textit{ Research exploring how to adapt existing specification and verification frameworks
for reactive systems to sequential DL models will be needed to broaden the class of DL systems that are 
amenable to verification and validation.}
\end{quote}

\section{Testing of Deep Learning Systems}
\label{sec:testing}

\subsection{Background}
\vspace{0.5em}

Testing DL applications using traditional SE infrastructure is hard. They often lack end-to-end formal specifications~\cite{taylor2005rule,seshia2016towards,dreossi2018semantic}. Manually creating a complete specification for complex systems like autonomous vehicles is hard, if not impossible, as it means mimicking all possible real-world scenarios. Researchers from Google Brain observed that while developing DL  applications, a developer ``had in mind a certain (perhaps informally specified) objective or task"; an {\em accident} occurs when the application ``that was designed for that task produced harmful and unexpected results''~\cite{Amodei}. At a conceptual level, these accidents are analogous to semantic bugs in traditional software. Similar to the bug detection and patching cycle in traditional software development, the erroneous behaviors of DNNs, once detected, can be fixed by either adding the error-inducing inputs to the training data set and/or changing the model structure/parameters.

However, a DL is very different from Traditional Software: while human developers manually write the core logic in the former and the logic is encoded in control and data flow, DLs learn their logic from a large amount of training data with minimal human guidance and the logic is encoded in terms of nonlinear activation functions and weights of edges between different neurons. These differences are particularly significant for software testing because testing essentially checks the program logic, which is encoded very differently in these two form of software. For example, code coverage guided testing techniques~\cite{hutchins1994experiments} that rely on control and data-flow logic will likely not work to test DNN logic~\cite{pei2017deepxplore}. Symbolic execution-based testing will also likely be difficult to adapt as it uses SMT solvers that known to have troubles with non-linearity~\cite{cadar2011symbolic}.

To this end, DL system-testing consists of addressing the following four challenges: (1) what components/properties of a DL system should be tested?; (2) how should inputs be generated to test them?; (3) how should progress be measured (akin to measuring testing effectiveness for "traditional software")?; and (4) how should debugging proceed when testing techniques uncover problems?

\subsection{Research Opportunities}
\vspace{0.5em}

\subsubsection{Determining What to Test}
\vspace{0.5em}

The goal of DL system testing is similar to that of exposing defects in traditional software systems. Therefore, the key question that must be answered is: \textit{what constitutes a DL system defect}. Such defects could be present in infrastructure (e.g., TensorFlow), DL application code, data distribution, model structure, weight values, and hyper-parameters. They may have various symptoms, including those similar to software 1.0 bugs such as exceptions/crashes, and others unique to the DL semantics, such as low model accuracy, difficulty in convergence, robustness issues, and malicious back-doors.  For some of these defects, the oracles (intended properties) can be explicitly defined, whereas for others, defining oracle is a prominent challenge. 

\begin{quote}
\textbf{T$_1$:}\textit{ Future work should focus on achieving a better understanding of the faults that might occur in DL-based systems through empirical work.}
\end{quote}

Many DL defects are rooted in low model accuracy. For example, many believe that inaccurate models tend to have robustness issues. Hence, a general test oracle may focus on model accuracy. The challenge is to factor in the discrepancy among data distributions during training, testing, and deployment. Metamorphic testing provides a potential solution by asserting model behaviors upon variations. It is also possible to use the software 1.0 version of the application or an interpretable approximation (e.g., decision tree) as the oracle to test a DL model. Specifically, many DL applications have their antecedents in traditional software, which is based on deterministic algorithms or rules. These algorithms and rules provide an approximation of the intended state space, allowing us to test DL models that are largely uninterpretable. When a DL model is potentially malicious, the properties to test may need to change. Low-level hygiene properties analogous to buffer bound checks in traditional software may need to be tested and validated.

\begin{quote}
\textbf{T$_2$:}\textit{ Researchers should focus on trying to draw analogies between testing practices that have been successfully applied to traditional software systems, and those adapting those to fit the needs of DL-based systems.}
\end{quote}

\subsubsection{Deciding how to Test DL-based Systems}
\vspace{0.5em}

Analogous to testing in traditional software, white-box, black-box, and grey-box testing techniques can be developed to test DL systems. White-box testing is driven by some coverage criteria. A number of such criteria have been proposed in the literature, such as neuron coverage~\cite{Tian:ICSE'18}, and have demonstrated potential in generating diverse inputs. In traditional software, various coverage criteria have different trade-offs in their cost and capabilities in disclosing software defects. For example, definition-use criterion that aim to cover all the dataflow relations in the subject software is much more expensive than statement coverage but much more effective in exposing bugs. Similar trade-offs exist in DL system white-box testing and hence studying their correlations with capabilities of disclosing model defects is of importance. Existing DL model coverage criteria mainly focus on specific model structures (e.g., CNN) and input modality. They can be extended to other structures such as RNN and other modalities. 

\begin{quote}
\textbf{T$_3$:}\textit{ There is a need for the development of proper test adequacy criteria for DL-based systems, akin to code coverage, that align with relevant abstractions of DL-based systems.}
\end{quote}

In traditional software, black-box testing is an important methodology in practice. It does not require access to software implementation or low level design documents. Instead, it directly derives test cases from functional specifications. Existing black-box DL system testing focuses on partitioning a pre-existing data set to training, validation, and test data sets and performs cross-validation. Data augmentation and GAN can be used to generate additional data. However, it is unclear if such data generation can be driven by model functional specifications. 

\begin{quote}
\textbf{T$_4$:}\textit{ Research on evolving practices for Black-box testing of DL systems needs to evolve to provide additional details to developers regarding model performance.}
\end{quote}

Recently, we have witnessed substantial progress in grey-box testing for traditional software systems. Numerous fuzzing techniques and search-based test generation techniques have advanced the state-of-the-art of traditional bug finding. It is likely that similar techniques can be developed to test DL systems. Model continuity and the presence of gradient information provide unique opportunities for such techniques. It is also foreseeable that many effective software 1.0 testing techniques such as mutation testing, unit testing, and regression testing will have their counter-parts in DL system testing. However, in software 1.0, mutating a program statement and testing a function/unit has clear semantics, the un-interpretability of DL models makes it difficult to associate clear meanings to mutating model weight values, model structures, and testing phases (in the pipeline)  and layers in models. There are many reasons to believe that differential testing provides unique benefits as it provides cross-referencing oracles, and leverages counter-factual causality to mitigate the inherent un-interpretability problem in DL systems. Most existing testing techniques focus on testing either DL models or non-model components (in traditional programming languages), co-testing them together as a cohesive system may pose unique challenges.

\begin{quote}
\textbf{T$_4$:}\textit{ There are many promising avenues of potential work on Grey-box testing of DL-systems, particularly related to techniques that take advantage of model continuity and gradient information to drive automated fuzzers or search-based techniques. Researchers should also look to find design counterparts to different types of traditional software testing (e.g. unit tests, regression tests) for DL-based systems.}
\end{quote}

\subsubsection{Determining How to Measure scientific progress on Testing DL-based Systems}
\vspace{0.5em}

Measuring progress is critical as testing is an iterative procedure that cannot expose all the bugs in the test subject. Hence, we need to know that sufficient progress has been made so that the procedure can be terminated. In software 1.0, error detection rate and coverage improvement are used to measure progress. We need to establish the counter-part in DL system testing. Measuring the improvement of model accuracy over time may not be sufficient as the training may fall into some local optima. Measuring model coverage (e.g., neuron coverage) is a plausible solution, although the correspondence between coverage and various model quality objectives needs to be established. Continuous testing after DL system is deployed is valuable as currently rigorous model testing and retraining (after deployment) only happen when things go very wrong. 

\begin{quote}
\textbf{T$_5$:}\textit{ Research into the design of test effectiveness metrics is important to measure the progress being made in testing research for DL-based systems.}
\end{quote}

\subsubsection{Determining How to Debug DL-based Systems}
\vspace{0.5em}

Once defects are disclosed, the subsequent challenge is how to fix these defects. The current practice in DL model engineering relies on trial and error, meaning that the data engineers make changes to various parts such as training data set, model structure and training parameters, which they believe will lead to defect mitigation based on their experience. It lacks a critical step of identifying the root causes of defects and using that as guidance to fix the problems. Effective diagnosis tools to help point to specific (defective) artifacts in the engineering pipeline and suggest possible fixes are hence of importance. Differential analysis that has been highly effective in diagnosing software 1.0 bugs could be valuable in model and infrastructure defect diagnosis. For example, eliminating features or model components and testing how the system performs could be one form of debugging/testing. In addition, while fixing software 1.0 defects largely lies in changing certain program statements, there is hardly a counter-part in fixing model defects. For example, directly changing some weight values have uninterpretable consequences on the model behaviors.

\begin{quote}
\textbf{T$_6$:}\textit{ Differential analysis may play a key role in aiding in debugging practices for DL-based systems, and near term research could benefit from building on such techniques.}
\end{quote}

\section{Development \& Deployment Challenges for Deep Learning Systems}
\label{sec:dev-deploy}

\subsection{Background}
\vspace{0.5em}

As DL continues to pervade modern software systems, developers and data scientist are beginning to grapple with development and deployment challenges that are markedly different from traditional processes. While the development process for traditional software systems has gone through many iterations (e.g., waterfall $\rightarrow$ agile), the steps in these process are generally fairly well-defined. However, for DL-based systems, the process is much more "experimental" in many regards. That is, developers must formulate hypotheses regarding their problem and dataset, create models, and "test" these models against their hypothesis to determine if they are capturing trends in the data effectively and making relevant predictions. Given the relatively experimental nature of the DL development process, there are many open questions regarding effective best-practices, tooling, and sociotechnical processes that represent promising areas for future work.

\subsection{Research Opportunities}
\vspace{0.5em}

\subsubsection{Requirements Engineering for DL Systems}
\vspace{0.5em}

Requirements engineering is a critical part of any software development process, as specifying what should be built can often be more difficult than the process of actually instantiating the ideas into code. However, the experimental nature of DL-based systems can make requirements more difficult to pin down. When used in industry, engineers often don't have formal requirements for a DL system~\cite{Amershi:ICSE'19}. This is due to the fact that, if a team is turning to a DL-based solution for a specific problem, this means that the problem domain is likely too complex to specify analytically. Therefore, teams often have a general goal in mind, and a specific success criteria for that goal. For most DL-based systems this goal is often exceeding a particular threshold for a given effectiveness metric drawn from ML literature, and when this threshold is met, further "optimal" performance is rarely sought after the fact. However, there are two main aspects of requirements engineering for DL systems that are potentially important and serve as areas of future work: (i) delineating and understanding the boundaries of performance of a given model, and (ii) non-functional requirements such as model size and inference time. In order to understand whether the predictive performance a given DL-model satisfies the requirements of a given problem, the limits and boundaries of these models must be explored and understood.

\begin{quote}
\textbf{DD$_1$:}\textit{ Research on developing techniques to properly specify the behavioral boundaries and limitations of DL-models will be an important aspect of requirements engineering to ensure proper effectiveness for a given problem domain.}
\end{quote}

In addition, given the problem domain, there be important non-functional requirements related to aspects such as model size, inference time, privacy and bias considerations, or memory size. Developers will require support in ensuring such requirements are met in practice.

\begin{quote}
\textbf{DD$_2$:}\textit{ Developers will need automated support for building DL-based systems that meet a variety of non-functional requirements including technical considerations such as model size, and non-technical considerations such as issues with privacy or bias.}
\end{quote}

In addition to requirements engineering, the notion of software traceability is also in its nascent stages for DL-based systems. In traditional software systems, the artifacts typically involved in the traceability process are usually well-defined, and often carry with them an inherent structure. However, this is not necessarily the case for DL-based systems. For instance, one could envision traceability tools that trace from data examples to different abstract data representations within a given model. The data-driven nature of DL also poses new challenges and opportunities for work on traceability. For example, given that understanding and cleaning data is such a large part of the DL development pipelines, traceability approaches could be reworked to offer links between various clusters of a given dataset and testing examples.

\begin{quote}
\textbf{DD$_3$:}\textit{ There is a need for traceability to be fundamentally rethought for DL-based systems. Researchers should focus on determining what types of trace links are necessary for DL-based systems and work toward automated approaches that can automatically infer and reason about such links.}
\end{quote}

\subsubsection{Sociotechnical Aspects of DL System Development}
\vspace{0.5em}

At its core, any type of software development is a process that is carried out by humans, and more often than not, groups of humans working together toward a common goal. Given this fact, the socio-technical processes of development are critical to successful creation and instantiation of software. With the rise of agile methodologies, and collaborative tools (e.g., issue trackers, software specific task managers) robust methods for cooperative development of traditional software systems have been developed. However, the experimental nature of DL-based systems and their differing iterative processes make it difficult to cleanly transfer many of the existing development processes and tools. 

\begin{quote}
\textbf{DD$_4$:}\textit{ Researchers and practitioners should work together to study and understand effective processes and tools for the development of DL systems. This will provide guidance for future work on more intelligent tools that improves or accelerates these processes.}
\end{quote}

In work that has examined the DL-like development processes carried out by data scientists~\cite{Amershi:ICSE'19}, it is clear that the most time- and effort-consuming portion of the development process is data curation and management. As the popularity of DL-systems have continued to improve, libraries and APIs for creating models have become much easier to use. However, these underlying models only function well if trained on properly curated datasets. Therefore, much of the engineering effort for such systems is focused on data. This is a fundamental shift from more traditional software development workflows, wherein the focus is largely on code. Instead, developers and data scientists must focus on \textit{understanding} data, and making sure that the data accurately reflects the problem that they are trying to solve, to extent that is possible. This means that various types of exploratory data analysis should be given the same levels of consideration as program analysis techniques have been given for code, if we are properly support these practitioners in their data-centric endeavors. 

\begin{quote}
\textbf{DD$_5$:}\textit{ Given the centrality of data in the development of DL-based systems, and the effort typically spent on data-related tasks, future research should focus on providing tools and techniques for data analysis that allow developers to better understand their datasets, and how the nature of their data will affect their DL-models, and vice-versa.}
\end{quote}

Once a DL system has been created, developers must then test their system to ensure that it is working as intended. While we discuss the various future directions of work for testing-DL based systems, there is one highly related process that is deeply intertwined with testing: \textit{debugging}. Once developers are able to test their systems, they must then attempt to remedy any detected misbehavior. In traditional software development, this typically means inspecting various aspects of the code, stepping through its execution, and determining failure cases. However, given the data-centric nature of DL systems, "debugging" the \textit{data} will be nearly as important as debugging model, and it is likely that the two tasks will be highly intertwined with one another. Thus, there are clear research opportunities for developing both processes and tools that aid developers in debugging DL-systems across both models and code.

\begin{quote}
\textbf{DD$_6$:}\textit{ There is a clear need for research into understanding and aiding in the debugging of DL-based systems. Such work will need to account for the data-driven nature of the systems and develop tools and techniques for debugging both models and data, and make considerations for collaborative debugging.}
\end{quote}

\subsubsection{Deployment and Monitoring of DL-based systems}
\vspace{0.5em}

Deployment and monitoring practices for traditional software systems have evolved markedly in recently years with advancements related both to processes and infrastructure supporting continuous integration (CI) and deployment (CD). However, it may difficult for such processes to be readily adapted for use in DL-based systems given the size of typical datasets and computational complexity of training DL models. For example, one could envision a CI system that retrains a series of models given updates to a tracked dataset. However, developer would likely need additional monitoring for training processes, and perhaps training processes would differ dynamically based iterative results during the process. Such support could prove difficult for current popular CI systems. Monitoring also poses unique challenges. While there could still exist field failures (akin to crashes) that signal issues with software, it is likely that unwarranted behavior will be more dependent upon end-users reporting anomalous behavior. Additionally, monitoring for performance-related metrics and making recommendations to assuage any potential issues is also likely to be important. These all challenges represent rich areas of research moving forward. 

\begin{quote}
\textbf{DD$_7$:}\textit{ Research should focus on how to adapt current practices for CI/CD to the context of DL-based systems, and develop effective monitoring solutions to capture field failures.}
\end{quote}

\subsubsection{Educational Aspects of the Development and Deployment of DL Systems}
\vspace{0.5em}

It is clear that there are many changes from the traditional software development process reflected in the comparable processes for DL-based systems. As research on these systems continues to evolve, and our understanding advances, the educational materials for students seeking careers in software development or data science must advance alongside it. 

\begin{quote}
\textbf{DD$_8$:}\textit{ As advancements in our understanding of effective development processes and techniques manifest, these must be reflected in freely available educational materials that prepare students accordingly. An academia-industry partnership for such materials would be beneficial to provide practical grounding for course materials.}
\end{quote}

\section{Maintenance of Deep Learning Systems}
\label{sec:maintenance}

\subsection{Background}
\vspace{0.5em}

Software maintenance is a key phase of the software development life-cycle wherein a system is modified to correct faults, add features, or to improve various other functional or non-functional properties. It has been estimated that nearly half of all software engineering activities are dedicated to one of the various types of maintenance activities~\cite{Lientz:1980}. As such, there has been a tremendous amount of research effort involved in analyzing and improving maintenance-related software engineering tasks. Such work has ranged from automated analysis of issue and bug reports~\cite{Moran:FSE'15,Chaparro:ICSME'16,Chaparro:ICSME'17,Moran:ICSE'16}, to full-fledged automated repair of faults in software systems~\cite{LeGoues-ICSE'12,Weimer:ICSE'09,Chen:TSE'19}.  However, until recently, much of this work has been focused primarily upon \textit{traditional} software systems written in an analytic nature. As noted at the beginning of this report, as software is applied to tackle increasingly complex tasks, there has been a shift from analytical development to \textit{learning-based} development, where machine learning algorithms are applied to large datasets to ``learn'' a program for a given computational task. The popularity of multi-layered Neural Networks and accompanying optimization algorithms (so-called Deep Learning architectures) have been a major driver of this phenomenon. However, such learning-based systems are inherently different from, and often intertwined with, more traditional software systems. Given the relative recency of DL-based software systems, there are many open questions regarding proper maintenance practices. We believe that developing and ensuring proper maintenance practices for DL-based systems is imperative, and we outline the key elements of a proposed research agenda below.

\subsection{Research Opportunities}
\vspace{0.5em}

\subsubsection{Measuring and Understanding the Evolution of Deep Learning Systems}
\vspace{0.5em}

The maintenance of a software system is tightly coupled to its evolution, i.e., the types and magnitude of maintenance performed and software artifacts present in a system often dictate how that system evolves over time. Different from more traditional software systems, DL systems have additional artifacts that must be accounted for during software evolution. For instance, such systems will typically consist of (i) model code that implements a given DL architecture, (ii) configuration files that specify different model hyper-parameters, (iii) trained models, (iv) datasets split for various use cases (\eg training/validation/test sets), and (v) performance or effectiveness metrics for different trained models. 
The process for constructing DL systems also fundamentally differs from more traditional software systems. DL-based systems are inherently more ``experimental'' in nature, wherein developers will construct, train, test, and tweak several different models and DL architecture configurations. Given the opacity of DL models, and the sometimes surprising nature of model results, engineers often implement a more exploratory process compared to more traditional software systems where the behaviors of analytical code are easier to predict. Given the differences in software artifacts and development pipelines between DL-based systems and traditional software systems, we would expect there to be distinct differences in their evolution and maintenance as well.

However, currently the software engineering research community is still grappling with the differences in process and evolutionary aspects of DL-based systems. We do not know how to precisely track
changes nor do we understand how they are typically manifested.

As indicated earlier, ML/DL-based systems are heavily data-centric from the viewpoint of developers~\cite{Amershi:ICSE'19}. That is, data is a crucial, and often unwieldy, component that enables DL architectures to learn ``programs'' for complex applications. However, the current state of data management for DL systems is a pain point for many developers, and thus should be a focus area for the research community. However, data is not the only artifact that must evolve over the lifespan of a DL-based software system. There are several interconnected artifacts such as trained models, model test results, hyper-parameter configurations, and model code that must all co-evolve in an efficient manner. The intermingling of these artifacts often results in a large amount of glue-code coupling these artifacts together. Designing both processes and techniques/tools to help measure  such co-evolution should be a focus of researchers moving forward to start taming this challenge.
\vspace{0.5em}

\begin{quote}
\textbf{M$_1$:}\textit{ Researchers should focus on designing techniques for efficient tracking and evolution of the rich and often tightly coupled software artifacts that are associated with DL-based systems.}
\end{quote}

A recent survey conducted at Microsoft has provided some insight into the processes by which ML/DL components are created and integrated into existing software systems~\cite{Amershi:ICSE'19}. This work has found three main aspects of ML/DL-based software systems that fundamentally differ from other domains including: (i) discovering, managing, and versioning data, (ii) model customization and reuse, and (iii) the integration of AI components into more traditional software systems. However, while this study begins to scratch the surface of evolutionary aspects of DL systems, there are still many open questions that remain, for instance: \textit{What does change look like in a DL-based system?}, \textit{How do models and datasets co-evolve?}, \textit{How important is it to maintain a version history for datasets?}, \textit{How does model code co-evolve with trained models and testing?}, \textit{Do different types of regressions befall DL-based systems?}. To answer such questions it is clear that additional empirical studies are needed to help guide eventual research towards supporting developers with some of the more challenging aspects of such evolution.

\begin{quote}
\textbf{M$_2$:}\textit{ Research empirically analyzing the evolutionary properties of DL-based systems will provide much needed direction towards understanding and eventually supporting developer needs throughout the software maintenance process.}
\end{quote}

\subsubsection{Grappling with Technical Debt and Maintenance in Deep Learning Systems}
\vspace{0.5em}

Due to the rapid pace of development typically associated with modern software systems, engineers are often seen as facing a tenuous dichotomy: move quickly and ship new features and completed projects or slow down to ensure the quality of engineering and the sound design and implementation of a system. This trade off between sound engineering practices and velocity of progress is often referred to as a concept called \textit{technical debt}. Technical debt shares several aspects in common with fiscal debt. For example, if poor engineering decisions are made for the sake of development speed and  are not corrected in a timely manner, this could result in increasing maintenance costs. Technical debt is an increasingly researched topic within the broader field of software engineering~\cite{Verdecchia:TechDebt'18}. However, currently there is only an early understanding regarding the types of technical debt incurred specifically by DL-based systems. Recent work from engineers and researchers at Google has shed light on the various forms of debt that more general ML systems can incur~\cite{Sculley:NIPS'14}. Some of the types of debt explored by this work include (i) the erosion of boundaries between software ML software components, (ii) System-level spaghetti, and (iii) changes in the external world (and representing data). While these types of debt generally apply to DL-based systems, this paper was a wider look at ML-based as a whole.  We are also witnessing the emergence of techniques such as refactoring, inspired by those in traditional software, but adapted to work on DL-based systems to address different types of technical debt \cite{shriver2019RefactoringNN}. 
There are likely to be even more  forms of technical debt that could be incurred by DL-based systems, for instance: \textit{What is the impact of a change to the data or the model?}, \textit{Are the changes to maintain accuracy diminishing robustness?}, and \textit{What is the effect of using an existing off-the shelf architecture on explainability?}.
Thus, there needs to be better understanding of the specific types of technical debt that DL systems may incur.

\begin{quote}
\textbf{M$_3$:}\textit{ Research into the trade-offs between development velocity and maintenance costs should be undertaken to empirically determine mechanisms to help developers manage such debt.}
\end{quote}

\subsubsection{Abstractions to enable Analyses}
\vspace{0.5em}

DL-based systems have unique computational encodings that render existing traditional  software abstractions largely obsolete. For instance, it is unclear how traditional control and data flow would map to  DL architectures  made up of several different computational layers that engineers may view as performing discrete functions. However, understanding these abstractions will be key towards enabling automated analysis of such systems in the future.

\begin{quote}
\textbf{M$_4$:}\textit{ Researchers should strive to understand the salient abstractions of DL-based systems, both mental and technical in hopes of better supporting analyses of DL-based systems.}
\end{quote}

Once a proper set of abstractions have been established, it is critical that work be conducted in order to aid in the automated analysis of DL-based programs. Decades of research on program analysis techniques of more traditional software systems have ushered in several advancements in automated developer tools and frameworks for software maintenance and validation. If we hope to achieve similar levels of advancement in the context of DL-based systems, researchers must work toward designing and building these next-generation analysis techniques. This could come in the form of instrumentation of dynamically running DL models, or via a combination of static analysis of model configuration and dynamic analysis. 

\begin{quote}
\textbf{M$_5$:}\textit{ Research into program analysis tools for DL-based systems is critical if the research community is to develop automated developer support for software maintenance tasks.}
\end{quote}

\subsubsection{Educational Challenges for Teaching Deep Learning System Maintenance}
\vspace{0.5em}

As with any emerging domain of software development, education is a critical component to ensuring engineers are equipped to work with DL-based systems in as effective and efficient a manner as possible. However, from a researchers perspective, it can be difficult to glean best-practices and in turn develop effective pedagogical mechanisms for conveying these to students. One potentially promising path forward for tackling the current gap in computer science education is for researchers/educators to collaborate with industry to learn current best practices and co-develop course materials and pedagogical techniques for conveying it.

\begin{quote}
\textbf{M$_6$:}\textit{ Researchers and Industrial practitioners should work together to identify effective development/maintenance practices for DL-based systems and co-design educational and pedagogical materials for conveying these to students.}
\end{quote}

\section{Deep Learning for Code Generation}
\label{sec:dlcodegen}

\subsection{Background}
\vspace{0.5em}

The task of program synthesis, i.e., automatically generating programs from specifications, is considered a fundamental problem in Artificial Intelligence. In recent years, there has been tremendous research progress in the field across various communities including Software Engineering, Programming Languages, and Artificial Intelligence. The advances in deep learning techniques coupled with neuro-symbolic reasoning techniques offer exciting opportunities to enable new automated code generation paradigms and interfaces for programmer-synthesizer collaboration.

\subsection{Research Opportunities}
\vspace{0.5em}

\subsubsection{Application Domains}
\vspace{0.5em}

There are a number of different application domains that might be well suited for automated code generation both in near term and longer term research. Some near term opportunities may lie in the areas of:
\begin{itemize}
    \item{\textbf{Program Superoptimization} -- automatically generating programs that are functionally equivalent to a given implementation but allow for complete transformation of original programs using search unlike traditional compilers.}
    \item{\textbf{Code completion} -- generating completions of small snippets of code given some code context.}
    \item{\textbf{Repairing programs with small program patches} -- Instead of generating complete code snippets, synthesis techniques can be used to generate small program patches that satisfy the failing specification or tests.}
    \item{\textbf{End-user programming} -- Helping millions of end-users who may not necessarily be programmers like spreadsheet users to accomplish programmatic tasks.}
    \item{\textbf{Mobile app development} -- Helping programmers develop mobile applications using natural language and examples.}
\end{itemize}

Another area of interest was for synthesizing programs in domain-specific languages such as SQL, yaml, and build files was identified as a promising opportunity. In these domains, programmers typically have knowledge about performing tasks in general purpose languages, but often have to look up the syntax and semantics of these domain-specific languages. There could be opportunities in considering natural language as intermediate representations for different computations and use that to transfer implementations in different domain-specific languages. One research challenge here would be on how to enable naturalness and readability of the automatically generated code for maintainability if developers use them as part of a larger workflow.

\begin{quote}
\textbf{DD$_1$:}\textit{ There a number of application domains that represent promising paths forward in DL for code generation including (i) program superoptimization, (ii) code completion, (iii) program repair, (iv) end-user programming, and (v) mobile app development. However, major challenges remain in ensuring that generate code is both easily comprehendable and maintainable.}
\end{quote}

A key challenge in program synthesis is that of specification, particularly for complex tasks. Even for simple programs, writing a full specification can sometimes be as tedious as writing the complete program in first place. There are several alternate specification mechanisms such as natural language, input-output examples, unit tests, partial implementations, existing implementations, program properties, and user interfaces, where different mechanisms are suited for different synthesis domains. One big opportunity with advances in deep learning techniques is to enable a rich environment that can embed truly multi-modal specifications in various forms as listed above. One particular specification mechanism that may be promising was to start with an existing implementation of code found with some keyword based search and learn to edit it in ways to satisfy the specification. Another important challenge to keep in mind here is that the expertise of users might also influence the types of specification mechanisms that are useful in practice -- e.g. inexperienced users might not be familiar with even good keywords.

\begin{quote}
\textbf{DD$_2$:}\textit{ Combining multi-modal DL models that are capable of mixing and taking advantage of different types of program specifications represent a promising path forward for research on code generation. However, researchers should also keep in mind the context of where the approach will be applied and ensure generation mechanism matches the user expertise.}
\end{quote}

Instead of fully automating the code generation process, we can also consider building an assistive agent for helping developers and focus more on the collaborative and creative aspects of development. Human developers are great at certain aspects of code development workflow and code generators would be good at certain other complementary aspects such as remembering large contexts. Combining the two sets of varying expertise would be an exciting research challenge. Proactive code completion is one such example, where a code assistant can recognize whether certain functionality is incomplete and suggest useful idioms or completions. Writing code is seldom a one-shot process even for human developers -- we usually iterate quite a few times before leading to a final implementation. One interesting challenge would be building a mixed-initiative dialog agent that switches between developer and code generator while asking questions to refine the intent and generating the desired implementation.

\begin{quote}
\textbf{DD$_3$:}\textit{ Researchers should carefully consider the interplay between a code generator and end-user developers, and develop mixed-initiative agents that are capable of working in concert with developers, rather than trying to automate end-to-end development tasks outright.}
\end{quote}

There is also a strong need for benchmarking the progress of the capability of program synthesizers as well as possibly developing some grand challenge problems for the community. One particular grand challenge was to build an intelligent synthesizer that can win a programming contest. Some other suggestions for grand challenge problems included building a synthesizer to pass an introductory programming class, enabling students to automatically generate mobile apps, and automatically answering questions on help forums such as StackOverflow. Evaluating synthesizers that require human interaction is also a complex question, and coming up with metrics to quantify and benchmarking such interactions would also be an interesting direction.

\begin{quote}
\textbf{DD$_4$:}\textit{ There is a need for the research community to develop comprehensive benchmarks or "grand challenges" for code generation to measure and spur progress (akin the imagenet challenge in computer vision research).}
\end{quote}

There are also many interesting research challenges in developing deep learning architectures that are specialized for embedding programs and specifications. Some recent models have shown that encoding structure and static analysis information about programs is useful for improving the generative process of programs. In addition, developing models that are capable of embedding dynamic and rich semantic information would be crucial for improving current program synthesis models.

\begin{quote}
\textbf{DD$_5$:}\textit{ There is a need for DL models that are able to specifically take advantage of the inherent static structure and dynamic properties of source code.}
\end{quote}

In summary, the advances in DL techniques offer exciting opportunities for building systems that can understand multi-modal specifications in various forms such as natural language, input-output examples and partial programs. Combining DL techniques with symbolic approaches can significantly improve the capabilities of current synthesizers by combining intuition-based reasoning with symbolic logical reasoning. This is an exciting area of research with many open challenges and fundamental problems, and can also be a pillar for advancing DL techniques.

\section{Workshop Structure}
\label{sec:workshop}

\subsection{Overview}

This report summarizes the findings in the NSF-sponsored "Workshop on Deep Learning \& Software Engineering", held in Sand Diego, California on November 10th - 12th, 2019. This was a community visioning workshop to identify academic and industrial research challenges and promising future directions for work that sit at the intersection of the disciplines of software engineering and deep learning. the workshop was organized by Denys Poshyvanyk, Baishakhi Ray, Kevin Moran, Prem Devanbu, Matthew Dwyer, Michael Lowry, Xiangyu Zhang, Rishabh Singh, and Sebastian Elbaum. In order to faciltate focused small discussion groups, both days of the workshop were closed to public, and the workshop organizers invited ~30 researchers across academia and industry to attend. The workshop was structured into three days. During the first day a series of lightning talks encompassing a range of topics at the intersection of DL \& SE were delivered by volunteers from the invited participants. The second day comprised the bulk of the workshop discussion and was organized into several Breakout Group Sessions with smaller groups discussing targeted topics, and Plenary Sessions where all participants convened to discuss the results of the Breakout Group discussions. The second day of the workshop was concluded with a Plenary Session that reflected on the most important aspects of the discussion throughout the day. On the third day of the workshop the organizing committee met to discuss the results of the workshop discussions and to begin drafting this report. We provide the workshop schedule below. This report was shared with all workshop participants before being released to wider community for open comments and discussion. The full workshop program is available at \url{https://dlse-workshop.gitlab.io/schedule/}.

\subsection{Community Contributions to the Report}

As indicated by the goals of the workshop, this workshop is meant to serve as a living document wherein community members contribute their views regarding promising research opportunities that sit at the intersection of Deep Learning \& Software Engineering. we encourage participation in contributing to this report by filing a Merge Request with the Report paper repository which is hosted on GitLab: 

\noindent\url{https://gitlab.com/dlse-workshop/dlse-workshop-community-report}.

\begin{quote}
\textbf{Sunday, November 10th 2019}
\end{quote}

\begin{itemize}
    \item{\textbf{5:00pm-6:00pm:} Dinner}\vspace{-0.5em}
    \item{\textbf{6:00pm-8:00pm:} Session 0: Lightning Talks}\vspace{-0.3em}
    \begin{itemize}
    \item{\textbf{6:00pm-6:30pm:} Lightning Talk 1: Debugging Deep Learning Models using Program Analysis Techniques}\vspace{-0.3em}
    \begin{itemize}
        \item{Speaker: Xiangyu Zhang - Purdue University}\vspace{0.3em}
    \end{itemize}
    \item{\textbf{6:30pm-7:00pm:} Lightning Talk 2: Neural Program Synthesis}\vspace{-0.3em}
    \begin{itemize}
        \item{Speaker: Rishabh Singh - Google} \vspace{0.3em}
    \end{itemize}
    \item{\textbf{7:00pm-7:30pm:} Lightning Talk 3: When Deep Learning Met Code Search}\vspace{-0.3em}
    \begin{itemize}
        \item{Speaker: Satish Chandra - Facebook} \vspace{0.3em}
    \end{itemize}
    \item{\textbf{7:30pm-8:00pm:} Lightning Talk 4: Connecting Natural Language and Code using Deep Learning}\vspace{-0.3em}
    \begin{itemize}
        \item{Speaker: Raymond Mooney - University of Texas at Austin} 
    \end{itemize}
\end{itemize}
\end{itemize}

\vspace{0.5em}
\begin{quote}
\textbf{Monday, November 11th 2019}
\end{quote}
    
\begin{itemize}
    \item{\textbf{8:30am-10:30am:} Session 1: Plenary - Workshop Introduction}\vspace{-0.5em}
    \begin{itemize}
        \item{\textbf{8:30pm-8:40pm:} Workshop Introduction}\vspace{-0.3em}
        \begin{itemize}
            \item{Speakers: Sol Greenspan, Denys Poshyvanyk, Baishakhi Ray}\vspace{0.3em}
        \end{itemize}
        \item{\textbf{8:45am-9:15am:} Overview of DL4SE and SE4DL}\vspace{-0.3em}
        \begin{itemize}
            \item{Speaker: Denys Poshyvanyk - William \& Mary} \vspace{0.3em}
        \end{itemize}
        \item{\textbf{9:15am-10:00am:} Participant Introductions}\vspace{-0.3em}
        \begin{itemize}
            \item{Speaker: All Participants} \vspace{0.3em}
        \end{itemize}
        \item{\textbf{10:00am-10:30am:} Introduction of Breakout Group Topics}\vspace{-0.3em}
        \begin{itemize}
            \item{Speakers: Session Leads} \vspace{0.3em}
        \end{itemize}
    \end{itemize}
    \item{\textbf{10:30am-10:45am:} Coffee Break}
    \item{\textbf{10:45am-1:00pm:} Session 2 - Breakout Sessions}
    \begin{itemize}
        \item{\textbf{10:45am-11:45pm:} Breakout Group 1: Deep Learning for Software Engineering}\vspace{-0.3em}
        \begin{itemize}
            \item{Group Leader: Prem Devanbu} \vspace{-0.3em}
             \item{Group Scribe: Denys Poshyvanyk} \vspace{0.3em}
        \end{itemize}
        \item{\textbf{10:45am-11:45am:} Breakout Group 2: Verification and Validation of Deep Learning Systems}\vspace{-0.3em}
        \begin{itemize}
            \item{Group Leader: Matthew Dwyer} \vspace{-0.3em}
             \item{Group Scribe: Sebastian Elbaum} \vspace{0.3em}
        \end{itemize}
        \item{\textbf{10:45am-11:45am:} Breakout Group 3: Development and Deployment Challenges for Deep Learning Systems}\vspace{-0.3em}
        \begin{itemize}
            \item{Group Leader: Mike Lowry} \vspace{-0.3em}
            \item{Group Scribe: Kevin Moran} \vspace{0.3em}
        \end{itemize}
        \item{\textbf{11:45am-1:00pm:} Plenary Discussion 1}\vspace{-0.3em}
        \begin{itemize}
            \item{Group Leader: Denys Poshyvanyk, Baishaki Ray} \vspace{-0.3em}
            \item{Group Scribe: Kevin Moran} \vspace{0.3em}
        \end{itemize}
    \end{itemize}
    \item{\textbf{1:00pm-2:00pm:} Lunch}
    \item{\textbf{2:00pm-3:45pm:} Session 3 - Breakout Sessions}
    \begin{itemize}
        \item{\textbf{2:00pm-3:00pm:} Breakout Group 4: Maintenance of Deep Learning Systems}\vspace{-0.3em}
        \begin{itemize}
            \item{Group Leader: Sebastian Elbaum} \vspace{-0.3em}
             \item{Group Scribe: Mike Lowry} \vspace{0.3em}
        \end{itemize}
        \item{\textbf{2:00pm-3:00pm:} Breakout Group 5: Testing of Deep Learning Systems}\vspace{-0.3em}
        \begin{itemize}
            \item{Group Leader: Xiangyu Zhang} \vspace{-0.3em}
             \item{Group Scribe: Matthew Dwyer} \vspace{0.3em}
        \end{itemize}
        \item{\textbf{2:00pm-3:00pm:} Breakout Group 6: Deep Learning for Code Generation}\vspace{-0.3em}
        \begin{itemize}
            \item{Group Leader: Rishabh Singh} \vspace{-0.3em}
            \item{Group Scribe: Kevin Moran} \vspace{0.3em}
        \end{itemize}
        \item{\textbf{3:00pm-3:45pm:} Plenary Discussion 2}\vspace{-0.3em}
        \begin{itemize}
            \item{Group Leader: Denys Poshyvanyk, Baishaki Ray} \vspace{-0.3em}
            \item{Group Scribe: Kevin Moran} \vspace{0.3em}
        \end{itemize}
    \end{itemize}
    \item{\textbf{3:45pm-4:00pm:} Lunch}
     \item{\textbf{3:45pm-4:00pm:} Session 3}
    \begin{itemize}
        \item{\textbf{4:00pm-4:30pm:} Continuation of Plenary Discussion 2}\vspace{-0.3em}
        \begin{itemize}
            \item{Group Leader: Denys Poshyvanyk, Baishaki Ray} \vspace{-0.3em}
            \item{Group Scribe: Kevin Moran} \vspace{0.3em}
        \end{itemize}
        \item{\textbf{10:45am-11:45pm:} Plenary Session 3: Cross-Cutting Concerns for DL \& SE}\vspace{-0.3em}
        \begin{itemize}
            \item{Group Leader: Denys Poshyvanyk} \vspace{-0.3em}
             \item{Group Scribe: Kevin Moran} \vspace{0.3em}
        \end{itemize}
    \end{itemize}
    \end{itemize}
    
\vspace{0.5em}
\begin{quote}
\textbf{Tuesday, November 12th 2019}
\end{quote}

\begin{itemize}
    \item{\textbf{8:30am-12:00pm:} Discussion \& Report Writing}\vspace{2em}
\end{itemize}

\newpage

\section{Community Comments}
\label{sec:comm-comments}

\subsection{Overview \& Purpose}

The purpose of this section is collect comments and additional points from the broader academic and industrial community conducting research that sits at the intersection of Deep Learning \& Software Engineering. Thus, each subsection listed below contains the comments from a different community member. If you would like to contribute, please visit the report GitLab repository (\url{https://gitlab.com/dlse-workshop/dlse-workshop-community-report}) and follow the contribution guidelines posted in the repository.

\newpage
\bibliographystyle{abbrv}
\bibliography{references}

\begin{thebibliography}{100}

\bibitem{smt-comp}
{14th International Satisfiability Modulo Theories Competition}.
\newblock https://smt-comp.github.io/2019/.

\bibitem{sv-comp}
{Competition on Software Verification}.
\newblock https://sv-comp.sosy-lab.org/2019/.

\bibitem{hw-comp}
{Hardware Model Checking Competition}.
\newblock http://fmv.jku.at/hwmcc19/index.html.

\bibitem{sat-comp}
{The International Satisfiability Competitions}.
\newblock http://www.satcompetition.org/.

\bibitem{Allamanis:SPLASH19}
M.~Allamanis.
\newblock The adverse effects of code duplication in machine learning models of
  code.
\newblock In {\em Proceedings of the 2019 ACM SIGPLAN International Symposium
  on New Ideas, New Paradigms, and Reflections on Programming and Software},
  Onward! 2019, page 143–153, New York, NY, USA, 2019. Association for
  Computing Machinery.

\bibitem{Amershi:ICSE'19}
S.~Amershi, A.~Begel, C.~Bird, R.~DeLine, H.~Gall, E.~Kamar, N.~Nagappan,
  B.~Nushi, and T.~Zimmermann.
\newblock Software engineering for machine learning: A case study.
\newblock In {\em Proceedings of the 41st International Conference on Software
  Engineering: Software Engineering in Practice}, ICSE-SEIP ’19, page
  291–300. IEEE Press, 2019.

\bibitem{Amodei}
D.~Amodei, C.~Olah, J.~Steinhardt, P.~Christiano, J.~S. Openai, D.~Man{\'{e}},
  and G.~Brain.
\newblock {Concrete Problems in AI Safety}.

\bibitem{ball2004slam}
T.~Ball, B.~Cook, V.~Levin, and S.~K. Rajamani.
\newblock Slam and static driver verifier: Technology transfer of formal
  methods inside microsoft.
\newblock In {\em International Conference on Integrated Formal Methods}, pages
  1--20. Springer, 2004.

\bibitem{barrett20136}
C.~Barrett, M.~Deters, L.~De~Moura, A.~Oliveras, and A.~Stump.
\newblock 6 years of smt-comp.
\newblock {\em Journal of Automated Reasoning}, 50(3):243--277, 2013.

\bibitem{barrett2010smt}
C.~Barrett, A.~Stump, and C.~Tinelli.
\newblock The smt-lib standard-version 2.0.
\newblock In {\em Proceedings of the 8th international workshop on
  satisfiability modulo theories, Edinburgh, Scotland,(SMT'10)}, 2010.

\bibitem{barrett-tinelli:2007:cvc3}
C.~Barrett and C.~Tinelli.
\newblock {CVC3}.
\newblock In {\em International Conference on Computer Aided Verification},
  pages 298--302, 2007.

\bibitem{Bastani:2016:MNN:3157382.3157391}
O.~Bastani, Y.~Ioannou, L.~Lampropoulos, D.~Vytiniotis, A.~V. Nori, and
  A.~Criminisi.
\newblock Measuring neural net robustness with constraints.
\newblock In {\em Proceedings of the 30th International Conference on Neural
  Information Processing Systems}, NIPS'16, pages 2621--2629, USA, 2016. Curran
  Associates Inc.

\bibitem{bastani2018verifiable}
O.~Bastani, Y.~Pu, and A.~Solar-Lezama.
\newblock Verifiable reinforcement learning via policy extraction.
\newblock In {\em Advances in neural information processing systems}, pages
  2494--2504, 2018.

\bibitem{bojarski-etal:corr:2016:DAVE2}
M.~Bojarski, D.~D. Testa, D.~Dworakowski, B.~Firner, B.~Flepp, P.~Goyal, L.~D.
  Jackel, M.~Monfort, U.~Muller, J.~Zhang, X.~Zhang, J.~Zhao, and K.~Zieba.
\newblock End to end learning for self-driving cars.
\newblock {\em CoRR}, abs/1604.07316, 2016.

\bibitem{Boopathy2019cnncert}
A.~Boopathy, T.-W. Weng, P.-Y. Chen, S.~Liu, and L.~Daniel.
\newblock Cnn-cert: An efficient framework for certifying robustness of
  convolutional neural networks.
\newblock In {\em AAAI}, Jan 2019.

\bibitem{DBLP:conf/nips/BunelTTKM18}
R.~R. Bunel, I.~Turkaslan, P.~H.~S. Torr, P.~Kohli, and P.~K. Mudigonda.
\newblock A unified view of piecewise linear neural network verification.
\newblock In {\em NeurIPS}, pages 4795--4804, 2018.

\bibitem{klee:osdi:2008}
C.~Cadar, D.~Dunbar, and D.~Engler.
\newblock {KLEE}: Unassisted and automatic generation of high-coverage tests
  for complex systems programs.
\newblock {\em Proc. 8th USENIX Symposium on Operating Systems Design and
  Implementation}, 2008.

\bibitem{cadar2008klee}
C.~Cadar, D.~Dunbar, and D.~R. Engler.
\newblock Klee: Unassisted and automatic generation of high-coverage tests for
  complex systems programs.
\newblock In {\em OSDI}, volume~8, pages 209--224, 2008.

\bibitem{cadar2011symbolic}
C.~Cadar, P.~Godefroid, S.~Khurshid, C.~S. Pasareanu, K.~Sen, N.~Tillmann, and
  W.~Visser.
\newblock Symbolic execution for software testing in practice: preliminary
  assessment.
\newblock In {\em Software Engineering (ICSE), 2011 33rd International
  Conference on}, pages 1066--1071. IEEE, 2011.

\bibitem{DBLP:journals/cacm/CadarS13}
C.~Cadar and K.~Sen.
\newblock Symbolic execution for software testing: three decades later.
\newblock {\em Commun. {ACM}}, 56(2):82--90, 2013.

\bibitem{Chaparro:ICSME'16}
O.~Chaparro, J.~M. Florez, and A.~Marcus.
\newblock On the {{Vocabulary Agreement}} in {{Software Issue Descriptions}}.
\newblock In {\em 2016 {{IEEE International Conference}} on {{Software
  Maintenance}} and {{Evolution}} ({{ICSME}})}, ICSME'16, pages 448--452, Oct.
  2016.
\newblock ISSN:.

\bibitem{Chaparro:ICSME'17}
O.~Chaparro, J.~M. Florez, and A.~Marcus.
\newblock Using {{Observed Behavior}} to {{Reformulate Queries}} during {{Text
  Retrieval}}-based {{Bug Localization}}.
\newblock In {\em 2017 {{IEEE International Conference}} on {{Software
  Maintenance}} and {{Evolution}} ({{ICSME}})}, ICSME'17, pages 376--387, Sept.
  2017.
\newblock ISSN:.

\bibitem{chapman2014we}
R.~Chapman and F.~Schanda.
\newblock Are we there yet? 20 years of industrial theorem proving with spark.
\newblock In {\em International Conference on Interactive Theorem Proving},
  pages 17--26. Springer, 2014.

\bibitem{Chen:TSE'19}
Z.~{Chen}, S.~J. {Kommrusch}, M.~{Tufano}, L.~{Pouchet}, D.~{Poshyvanyk}, and
  M.~{Monperrus}.
\newblock Sequencer: Sequence-to-sequence learning for end-to-end program
  repair.
\newblock {\em IEEE Transactions on Software Engineering}, pages 1--1, 2019.

\bibitem{clarke2000counterexample}
E.~Clarke, O.~Grumberg, S.~Jha, Y.~Lu, and H.~Veith.
\newblock Counterexample-guided abstraction refinement.
\newblock In {\em International Conference on Computer Aided Verification},
  pages 154--169. Springer, 2000.

\bibitem{clarke-grumberg-peled:1999:book}
E.~M. Clarke, Jr., O.~Grumberg, and D.~A. Peled.
\newblock {\em Model Checking}.
\newblock MIT Press, 1999.

\bibitem{clarke1976system}
L.~Clarke.
\newblock A system to generate test data and symbolically execute programs.
\newblock {\em Software Engineering, IEEE Transactions on}, 3(1):215--222,
  1976.

\bibitem{7332513}
C.~S. {Corley}, K.~{Damevski}, and N.~A. {Kraft}.
\newblock Exploring the use of deep learning for feature location.
\newblock In {\em 2015 IEEE International Conference on Software Maintenance
  and Evolution (ICSME)}, pages 556--560, Sep. 2015.

\bibitem{crocker2007verification}
D.~Crocker and J.~Carlton.
\newblock Verification of c programs using automated reasoning.
\newblock In {\em Fifth IEEE International Conference on Software Engineering
  and Formal Methods (SEFM 2007)}, pages 7--14. IEEE, 2007.

\bibitem{das2002esp}
M.~Das, S.~Lerner, and M.~Seigle.
\newblock Esp: Path-sensitive program verification in polynomial time.
\newblock In {\em Proceedings of the ACM SIGPLAN 2002 Conference on Programming
  language design and implementation}, pages 57--68, 2002.

\bibitem{de:2008:z3}
L.~De~Moura and N.~Bj{\o}rner.
\newblock Z3: An efficient {SMT} solver.
\newblock In {\em Tools and Algorithms for the Construction and Analysis of
  Systems}, pages 337--340, 2008.

\bibitem{dreossi2018semantic}
T.~Dreossi, S.~Jha, and S.~A. Seshia.
\newblock Semantic adversarial deep learning.
\newblock In {\em International Conference on Computer-Aided Verification
  (CAV)}, 2018.

\bibitem{DBLP:conf/nfm/DuttaJST18}
S.~Dutta, S.~Jha, S.~Sankaranarayanan, and A.~Tiwari.
\newblock Output range analysis for deep feedforward neural networks.
\newblock In {\em {NFM}}, volume 10811 of {\em Lecture Notes in Computer
  Science}, pages 121--138. Springer, 2018.

\bibitem{Dvijotham18}
K.~Dvijotham, R.~Stanforth, S.~Gowal, T.~Mann, and P.~Kohli.
\newblock A dual approach to scalable verification of deep networks.
\newblock In {\em Proceedings of the Thirty-Fourth Conference Annual Conference
  on Uncertainty in Artificial Intelligence (UAI-18)}, pages 162--171,
  Corvallis, Oregon, 2018. AUAI Press.

\bibitem{DBLP:conf/atva/Ehlers17}
R.~Ehlers.
\newblock Formal verification of piece-wise linear feed-forward neural
  networks.
\newblock In {\em {ATVA}}, volume 10482 of {\em Lecture Notes in Computer
  Science}, pages 269--286. Springer, 2017.

\bibitem{afl}
M.~Z. et~al.
\newblock American fuzzy lop -- a security-oriented fuzzer.
\newblock {https://github.com/google/AFL}, 2020.

\bibitem{Fu:FSE'17}
W.~Fu and T.~Menzies.
\newblock Easy over {{Hard}}: {{A Case Study}} on {{Deep Learning}}.
\newblock In {\em Proceedings of the 2017 11th {{Joint Meeting}} on
  {{Foundations}} of {{Software Engineering}}}, FSE'17, pages 49--60,
  Paderborn, Germany, 2017. {ACM}.

\bibitem{ai2}
T.~{Gehr}, M.~{Mirman}, D.~{Drachsler-Cohen}, P.~{Tsankov}, S.~{Chaudhuri}, and
  M.~{Vechev}.
\newblock Ai2: Safety and robustness certification of neural networks with
  abstract interpretation.
\newblock In {\em 2018 IEEE Symposium on Security and Privacy (SP)}, pages
  3--18, May 2018.

\bibitem{goodfellow2014explaining}
I.~Goodfellow, J.~Shlens, and C.~Szegedy.
\newblock {Explaining and Harnessing Adversarial Examples}.
\newblock In {\em International Conference on Learning Representations (ICLR)},
  2015.

\bibitem{gopinathASE19}
D.~Gopinath, H.~Converse, C.~S. Pasareanu, and A.~Taly.
\newblock Property inference for deep neural networks.
\newblock In {\em 34th {IEEE/ACM} International Conference on Automated
  Software Engineering, {ASE} 2019, San Diego, CA, USA, November 11-15, 2019},
  pages 797--809, 2019.

\bibitem{DBLP:conf/icse/GopinathPWZK19}
D.~Gopinath, C.~S. Pasareanu, K.~Wang, M.~Zhang, and S.~Khurshid.
\newblock Symbolic execution for attribution and attack synthesis in neural
  networks.
\newblock In {\em Proceedings of the 41st International Conference on Software
  Engineering: Companion Proceedings, {ICSE} 2019, Montreal, QC, Canada, May
  25-31, 2019}, pages 282--283, 2019.

\bibitem{Gu:2018:DCS:3180155.3180167}
X.~Gu, H.~Zhang, and S.~Kim.
\newblock Deep code search.
\newblock In {\em Proceedings of the 40th International Conference on Software
  Engineering}, ICSE '18, pages 933--944, New York, NY, USA, 2018. ACM.

\bibitem{7985645}
J.~{Guo}, J.~{Cheng}, and J.~{Cleland-Huang}.
\newblock Semantically enhanced software traceability using deep learning
  techniques.
\newblock In {\em 2017 IEEE/ACM 39th International Conference on Software
  Engineering (ICSE)}, pages 3--14, May 2017.

\bibitem{holzmann2018power}
G.~J. Holzmann.
\newblock The power of ten--rules for developing safety critical code1.
\newblock {\em Software Technology: 10 Years of Innovation in IEEE Computer},
  2018.

\bibitem{huang:2017:CAV}
X.~Huang, M.~Kwiatkowska, S.~Wang, and M.~Wu.
\newblock Safety verification of deep neural networks.
\newblock In {\em Computer Aided Verification - 29th International Conference,
  {CAV} 2017, Heidelberg, Germany, July 24-28, 2017, Proceedings, Part {I}},
  pages 3--29, 2017.

\bibitem{hutchins1994experiments}
M.~Hutchins, H.~Foster, T.~Goradia, and T.~Ostrand.
\newblock Experiments of the effectiveness of dataflow-and controlflow-based
  test adequacy criteria.
\newblock In {\em Proceedings of the 16th international conference on Software
  engineering}, pages 191--200. IEEE Computer Society Press, 1994.

\bibitem{katz-etal:CAV:2017}
G.~Katz, C.~W. Barrett, D.~L. Dill, K.~Julian, and M.~J. Kochenderfer.
\newblock Reluplex: An efficient {SMT} solver for verifying deep neural
  networks.
\newblock In {\em Computer Aided Verification - 29th International Conference,
  {CAV} 2017, Heidelberg, Germany, July 24-28, 2017, Proceedings, Part {I}},
  pages 97--117, 2017.

\bibitem{katz2019marabou}
G.~Katz, D.~A. Huang, D.~Ibeling, K.~Julian, C.~Lazarus, R.~Lim, P.~Shah,
  S.~Thakoor, H.~Wu, and A.~Zelji{\'c}.
\newblock The marabou framework for verification and analysis of deep neural
  networks.
\newblock In {\em International Conference on Computer Aided Verification},
  pages 443--452. Springer, 2019.

\bibitem{khurshid-pasareanu-visser:2003:tacas}
S.~Khurshid, C.~S. P\v{a}s\v{a}reanu, and W.~Visser.
\newblock Generalized symbolic execution for model checking and testing.
\newblock In {\em Tools and Algorithms for the Construction and Analysis of
  Systems}, pages 553--568, 2003.

\bibitem{king1976symbolic}
J.~C. King.
\newblock Symbolic execution and program testing.
\newblock {\em Communications of the ACM}, 19(7):385--394, 1976.

\bibitem{LeGoues-ICSE'12}
C.~Le~Goues, M.~Dewey-Vogt, S.~Forrest, and W.~Weimer.
\newblock A systematic study of automated program repair: Fixing 55 out of 105
  bugs for \$8 each.
\newblock In {\em Proceedings of the 34th International Conference on Software
  Engineering}, ICSE ’12, page 3–13. IEEE Press, 2012.

\bibitem{leavens1999jml}
G.~T. Leavens, A.~L. Baker, and C.~Ruby.
\newblock Jml: A notation for detailed design.
\newblock In {\em Behavioral specifications of Businesses and Systems}, pages
  175--188. Springer, 1999.

\bibitem{LeClair:ICSE'19}
A.~LeClair, S.~Jiang, and C.~McMillan.
\newblock A neural model for generating natural language summaries of program
  subroutines.
\newblock In {\em Proceedings of the 41st International Conference on Software
  Engineering}, ICSE ’19, page 795–806. IEEE Press, 2019.

\bibitem{8433204}
D.~{Li}, Z.~{Wang}, and Y.~{Xue}.
\newblock Fine-grained android malware detection based on deep learning.
\newblock In {\em 2018 IEEE Conference on Communications and Network Security
  (CNS)}, pages 1--2, May 2018.

\bibitem{Lientz:1980}
B.~P. Lientz and E.~B. Swanson.
\newblock {\em Software Maintenance Management}.
\newblock Addison-Wesley Longman Publishing Co., Inc., USA, 1980.

\bibitem{dnnverificationsurvey}
C.~Liu, T.~Arnon, C.~Lazarus, C.~Barrett, and M.~J. Kochenderfer.
\newblock Algorithms for verifying deep neural networks, 2019.

\bibitem{liu2013effectively}
H.~Liu, F.-C. Kuo, D.~Towey, and T.~Y. Chen.
\newblock How effectively does metamorphic testing alleviate the oracle
  problem?
\newblock {\em IEEE Transactions on Software Engineering}, 40(1):4--22, 2013.

\bibitem{Liu:2018:DLB:3238147.3238166}
H.~Liu, Z.~Xu, and Y.~Zou.
\newblock Deep learning based feature envy detection.
\newblock In {\em Proceedings of the 33rd ACM/IEEE International Conference on
  Automated Software Engineering}, ASE 2018, pages 385--396, New York, NY, USA,
  2018. ACM.

\bibitem{7985701}
P.~{Liu}, X.~{Zhang}, M.~{Pistoia}, Y.~{Zheng}, M.~{Marques}, and L.~{Zeng}.
\newblock Automatic text input generation for mobile testing.
\newblock In {\em 2017 IEEE/ACM 39th International Conference on Software
  Engineering (ICSE)}, pages 643--653, May 2017.

\bibitem{loquercio-etal:RAL:2018:dronet}
A.~Loquercio, A.~I. Maqueda, C.~R.~D. Blanco, and D.~Scaramuzza.
\newblock Dronet: Learning to fly by driving.
\newblock {\em {IEEE} Robotics and Automation Letters}, 2018.

\bibitem{Ma:2018:DMT:3238147.3238202}
L.~Ma, F.~Juefei-Xu, F.~Zhang, J.~Sun, M.~Xue, B.~Li, C.~Chen, T.~Su, L.~Li,
  Y.~Liu, J.~Zhao, and Y.~Wang.
\newblock Deepgauge: Multi-granularity testing criteria for deep learning
  systems.
\newblock In {\em Proceedings of the 33rd ACM/IEEE International Conference on
  Automated Software Engineering}, ASE 2018, pages 120--131, New York, NY, USA,
  2018. ACM.

\bibitem{mannaTLbook}
Z.~Manna and A.~Pnueli.
\newblock {\em The temporal logic of reactive and concurrent systems -
  specification}.
\newblock Springer, 1992.

\bibitem{meyer1992applying}
B.~Meyer.
\newblock Applying'design by contract'.
\newblock {\em Computer}, 25(10):40--51, 1992.

\bibitem{Moran:FSE'15}
K.~Moran, M.~Linares-V{\'a}squez, C.~Bernal-C{\'a}rdenas, and D.~Poshyvanyk.
\newblock Auto-completing {{Bug Reports}} for {{Android Applications}}.
\newblock In {\em Proceedings of the 2015 10th {{Joint Meeting}} on
  {{Foundations}} of {{Software Engineering}}}, FSE'15, pages 673--686,
  Bergamo, Italy, 2015. {ACM}.

\bibitem{Moran:ICSE'16}
K.~Moran, M.~Linares-V{\'a}squez, C.~Bernal-C{\'a}rdenas, and D.~Poshyvanyk.
\newblock {{FUSION}}: {{A Tool}} for {{Facilitating}} and {{Augmenting Android
  Bug Reporting}}.
\newblock In {\em {{ICSE}}'16}, ICSE'16, May 2016.

\bibitem{Moran:TSE'18}
K.~P. Moran, C.~Bernal-Cárdenas, M.~Curcio, R.~Bonett, and D.~Poshyvanyk.
\newblock Machine learning-based prototyping of graphical user interfaces for
  mobile apps.
\newblock {\em IEEE Transactions on Software Engineering}, pages 1--1, 2018.

\bibitem{murphy2008properties}
C.~Murphy, G.~E. Kaiser, L.~Hu, and L.~Wu.
\newblock Properties of machine learning applications for use in metamorphic
  testing.
\newblock In {\em Proceedings of the Twentieth International Conference on
  Software Engineering {\&} Knowledge Engineering (SEKE'2008), San Francisco,
  CA, USA, July 1-3, 2008}, pages 867--872, 2008.

\bibitem{nielson-etal:2005:principles-of-pa}
F.~Nielson, H.~R. Nielson, and C.~Hankin.
\newblock {\em Principles of Program Analysis}.
\newblock Springer, 2005.

\bibitem{DBLP:conf/icml/OdenaOAG19}
A.~Odena, C.~Olsson, D.~Andersen, and I.~J. Goodfellow.
\newblock Tensorfuzz: Debugging neural networks with coverage-guided fuzzing.
\newblock In {\em Proceedings of the 36th International Conference on Machine
  Learning, {ICML} 2019, 9-15 June 2019, Long Beach, California, {USA}}, pages
  4901--4911, 2019.

\bibitem{papernot2016practical}
N.~Papernot, P.~McDaniel, I.~Goodfellow, S.~Jha, Z.~Berkay~Celik, and A.~Swami.
\newblock {Practical Black-Box Attacks against Deep Learning Systems using
  Adversarial Examples}.
\newblock In {\em ACM Asia Conference on Computer and Communications Security},
  2017.

\bibitem{bleu}
K.~Papineni, S.~Roukos, T.~Ward, and W.~jing Zhu.
\newblock Bleu: a method for automatic evaluation of machine translation.
\newblock pages 311--318, 2002.

\bibitem{pei2017deepxplore}
K.~Pei, Y.~Cao, J.~Yang, and S.~Jana.
\newblock Deepxplore: Automated whitebox testing of deep learning systems.
\newblock {\em arXiv preprint arXiv:1705.06640}, 2017.

\bibitem{DBLP:conf/iclr/RaghunathanSL18}
A.~Raghunathan, J.~Steinhardt, and P.~Liang.
\newblock Certified defenses against adversarial examples.
\newblock In {\em {ICLR}}. OpenReview.net, 2018.

\bibitem{Robbes:ICPC07}
R.~{Robbes} and M.~{Lanza}.
\newblock Characterizing and understanding development sessions.
\newblock In {\em 15th IEEE International Conference on Program Comprehension
  (ICPC '07)}, pages 155--166, 2007.

\bibitem{rosenblum1995practical}
D.~S. Rosenblum.
\newblock A practical approach to programming with assertions.
\newblock {\em IEEE Transactions on software engineering}, 21(1):19--31, 1995.

\bibitem{DBLP:conf/ijcai/RuanHK18}
W.~Ruan, X.~Huang, and M.~Kwiatkowska.
\newblock Reachability analysis of deep neural networks with provable
  guarantees.
\newblock In {\em {IJCAI}}, pages 2651--2659. ijcai.org, 2018.

\bibitem{Sculley:NIPS'14}
D.~Sculley, G.~Holt, D.~Golovin, E.~Davydov, T.~Phillips, D.~Ebner,
  V.~Chaudhary, and M.~Young.
\newblock Machine learning: The high interest credit card of technical debt.
\newblock In {\em SE4ML: Software Engineering for Machine Learning (NIPS 2014
  Workshop)}, 2014.

\bibitem{sen-etal:2005:cute}
K.~Sen, D.~Marinov, and G.~Agha.
\newblock {CUTE:} a concolic unit testing engine for {C}.
\newblock In {\em Proceedings of the 10th European Software Engineering
  Conference held jointly with 13th {ACM} {SIGSOFT} International Symposium on
  Foundations of Software Engineering, 2005, Lisbon, Portugal, September 5-9,
  2005}, pages 263--272, 2005.

\bibitem{seshia2016towards}
S.~A. Seshia, D.~Sadigh, and S.~S. Sastry.
\newblock Towards verified artificial intelligence.
\newblock {\em arXiv preprint arXiv:1606.08514}, 2016.

\bibitem{setiono1995understanding}
R.~Setiono and H.~Liu.
\newblock Understanding neural networks via rule extraction.
\newblock In {\em IJCAI}, volume~1, pages 480--485, 1995.

\bibitem{shriver2019RefactoringNN}
D.~Shriver, D.~Xu, S.~G. Elbaum, and M.~B. Dwyer.
\newblock Refactoring neural networks for verification.
\newblock {\em ArXiv}, abs/1908.08026, 2019.

\bibitem{NIPS2018_8278}
G.~Singh, T.~Gehr, M.~Mirman, M.~P\"{u}schel, and M.~Vechev.
\newblock Fast and effective robustness certification.
\newblock In S.~Bengio, H.~Wallach, H.~Larochelle, K.~Grauman, N.~Cesa-Bianchi,
  and R.~Garnett, editors, {\em Advances in Neural Information Processing
  Systems 31}, pages 10802--10813. Curran Associates, Inc., 2018.

\bibitem{singh-etal:POPL:2019:deeppoly}
G.~Singh, T.~Gehr, M.~P{\"{u}}schel, and M.~T. Vechev.
\newblock An abstract domain for certifying neural networks.
\newblock {\em {PACMPL}}, 3({POPL}):41:1--41:30, 2019.

\bibitem{sun-etal:ASE:2018}
Y.~Sun, M.~Wu, W.~Ruan, X.~Huang, M.~Kwiatkowska, and D.~Kroening.
\newblock Concolic testing for deep neural networks.
\newblock In {\em Automated Software Engineering (ASE)}, pages 109--119. ACM,
  2018.

\bibitem{tp-bench}
G.~Sutcliffe and C.~Suttner.
\newblock The tptp problem library.
\newblock {\em Journal of Automated Reasoning}, 21(2):177--203, 1998.

\bibitem{szegedy2013intriguing}
C.~Szegedy, W.~Zaremba, I.~Sutskever, J.~Bruna, D.~Erhan, I.~Goodfellow, and
  R.~Fergus.
\newblock Intriguing properties of neural networks.
\newblock In {\em International Conference on Learning Representations (ICLR)},
  2014.

\bibitem{taylor2005rule}
B.~J. Taylor and M.~A. Darrah.
\newblock Rule extraction as a formal method for the verification and
  validation of neural networks.
\newblock In {\em Neural Networks, 2005. IJCNN'05. Proceedings. 2005 IEEE
  International Joint Conference on}, volume~5, pages 2915--2920. IEEE, 2005.

\bibitem{8453089}
Y.~{Tian}, K.~{Pei}, S.~{Jana}, and B.~{Ray}.
\newblock Deeptest: Automated testing of deep-neural-network-driven autonomous
  cars.
\newblock In {\em 2018 IEEE/ACM 40th International Conference on Software
  Engineering (ICSE)}, pages 303--314, May 2018.

\bibitem{Tian:ICSE'18}
Y.~Tian, K.~Pei, S.~Jana, and B.~Ray.
\newblock Deeptest: Automated testing of deep-neural-network-driven autonomous
  cars.
\newblock In {\em Proceedings of the 40th International Conference on Software
  Engineering}, ICSE ’18, page 303–314, New York, NY, USA, 2018.
  Association for Computing Machinery.

\bibitem{tjeng2018evaluating}
V.~Tjeng, K.~Y. Xiao, and R.~Tedrake.
\newblock Evaluating robustness of neural networks with mixed integer
  programming.
\newblock In {\em International Conference on Learning Representations}, 2019.

\bibitem{Tufano:2018:EIL:3238147.3240732}
M.~Tufano, C.~Watson, G.~Bavota, M.~Di~Penta, M.~White, and D.~Poshyvanyk.
\newblock An empirical investigation into learning bug-fixing patches in the
  wild via neural machine translation.
\newblock In {\em Proceedings of the 33rd ACM/IEEE International Conference on
  Automated Software Engineering}, ASE 2018, pages 832--837, New York, NY, USA,
  2018. ACM.

\bibitem{Verdecchia:TechDebt'18}
R.~Verdecchia, I.~Malavolta, and P.~Lago.
\newblock Architectural technical debt identification: The research landscape.
\newblock In {\em Proceedings of the 2018 International Conference on Technical
  Debt}, TechDebt ’18, page 11–20, New York, NY, USA, 2018. Association for
  Computing Machinery.

\bibitem{DBLP:conf/icml/VermaMSKC18}
A.~Verma, V.~Murali, R.~Singh, P.~Kohli, and S.~Chaudhuri.
\newblock Programmatically interpretable reinforcement learning.
\newblock In {\em Proceedings of the 35th International Conference on Machine
  Learning, {ICML} 2018, Stockholmsm{\"{a}}ssan, Stockholm, Sweden, July 10-15,
  2018}, pages 5052--5061, 2018.

\bibitem{Wan:2018:IAS:3238147.3238206}
Y.~Wan, Z.~Zhao, M.~Yang, G.~Xu, H.~Ying, J.~Wu, and P.~S. Yu.
\newblock Improving automatic source code summarization via deep reinforcement
  learning.
\newblock In {\em Proceedings of the 33rd ACM/IEEE International Conference on
  Automated Software Engineering}, ASE 2018, pages 397--407, New York, NY, USA,
  2018. ACM.

\bibitem{7886912}
S.~{Wang}, T.~{Liu}, and L.~{Tan}.
\newblock Automatically learning semantic features for defect prediction.
\newblock In {\em 2016 IEEE/ACM 38th International Conference on Software
  Engineering (ICSE)}, pages 297--308, May 2016.

\bibitem{DBLP:conf/nips/WangPWYJ18}
S.~Wang, K.~Pei, J.~Whitehouse, J.~Yang, and S.~Jana.
\newblock Efficient formal safety analysis of neural networks.
\newblock In {\em NeurIPS}, pages 6369--6379, 2018.

\bibitem{DBLP:conf/uss/WangPWYJ18}
S.~Wang, K.~Pei, J.~Whitehouse, J.~Yang, and S.~Jana.
\newblock Formal security analysis of neural networks using symbolic intervals.
\newblock In {\em {USENIX} Security Symposium}, pages 1599--1614. {USENIX}
  Association, 2018.

\bibitem{Weimer:ICSE'09}
W.~Weimer, T.~Nguyen, C.~Le~Goues, and S.~Forrest.
\newblock Automatically finding patches using genetic programming.
\newblock In {\em Proceedings of the 31st International Conference on Software
  Engineering}, ICSE ’09, page 364–374, USA, 2009. IEEE Computer Society.

\bibitem{DBLP:conf/icml/WengZCSHDBD18}
T.~Weng, H.~Zhang, H.~Chen, Z.~Song, C.~Hsieh, L.~Daniel, D.~S. Boning, and
  I.~S. Dhillon.
\newblock Towards fast computation of certified robustness for relu networks.
\newblock In {\em {ICML}}, volume~80 of {\em Proceedings of Machine Learning
  Research}, pages 5273--5282. {PMLR}, 2018.

\bibitem{White:2016:DLC:2970276.2970326}
M.~White, M.~Tufano, C.~Vendome, and D.~Poshyvanyk.
\newblock Deep learning code fragments for code clone detection.
\newblock In {\em Proceedings of the 31st IEEE/ACM International Conference on
  Automated Software Engineering}, ASE 2016, pages 87--98, New York, NY, USA,
  2016. ACM.

\bibitem{DBLP:conf/icml/WongK18}
E.~Wong and J.~Z. Kolter.
\newblock Provable defenses against adversarial examples via the convex outer
  adversarial polytope.
\newblock In {\em {ICML}}, volume~80 of {\em Proceedings of Machine Learning
  Research}, pages 5283--5292. {PMLR}, 2018.

\bibitem{8318388}
W.~{Xiang}, H.~{Tran}, and T.~T. {Johnson}.
\newblock Output reachable set estimation and verification for multilayer
  neural networks.
\newblock {\em IEEE Transactions on Neural Networks and Learning Systems},
  29(11):5777--5783, Nov 2018.

\bibitem{xie2019coverage}
X.~Xie, H.~Chen, Y.~Li, L.~Ma, Y.~Liu, and J.~Zhao.
\newblock Coverage-guided fuzzing for feedforward neural networks.
\newblock In {\em 2019 34th IEEE/ACM International Conference on Automated
  Software Engineering (ASE)}, pages 1162--1165. IEEE, 2019.

\bibitem{xie2019deephunter}
X.~Xie, L.~Ma, F.~Juefei-Xu, M.~Xue, H.~Chen, Y.~Liu, J.~Zhao, B.~Li, J.~Yin,
  and S.~See.
\newblock Deephunter: a coverage-guided fuzz testing framework for deep neural
  networks.
\newblock In {\em Proceedings of the 28th ACM SIGSOFT International Symposium
  on Software Testing and Analysis}, pages 146--157, 2019.

\bibitem{Xu:2016:PSL:2970276.2970357}
B.~Xu, D.~Ye, Z.~Xing, X.~Xia, G.~Chen, and S.~Li.
\newblock Predicting semantically linkable knowledge in developer online forums
  via convolutional neural network.
\newblock In {\em Proceedings of the 31st IEEE/ACM International Conference on
  Automated Software Engineering}, ASE 2016, pages 51--62, New York, NY, USA,
  2016. ACM.

\end{thebibliography}



\end{document}